\documentclass[]{aastex631}

\newcommand{\fermi}{\mbox{\textit{Fermi}-GBM~}}
\newcommand{\ferminosp}{\mbox{\textit{Fermi}-GBM}}

\usepackage{comment, soul}
\usepackage{amsmath}
\usepackage{multirow}

\graphicspath{{./}{figures/}}

\shorttitle{11-year GBM Transient Search}
\shortauthors{Y. Kaneko}

\accepted{for publication in \textit{ApJS}}

\begin{document}

\title{An 11-Year Catalog of Gamma-Ray Transients: A Comprehensive Search with \textit{Fermi} Gamma-ray Burst Monitor Data}

\correspondingauthor{Yuki Kaneko}
\email{yuki@sabanciuniv.edu}

\author[0000-0002-1861-5703]{Yuki Kaneko}
\affiliation{Sabanc\i~University, Faculty of Engineering and Natural Sciences, \.Istanbul 34956 Turkey}
\author[0000-0001-9711-4343]{\"Ozge Keskin}
\affiliation{Sabanc\i~University, Faculty of Engineering and Natural Sciences, \.Istanbul 34956 Turkey}
\author[0000-0003-3791-3754]{Can G\"ung\"or}
\affiliation{\.Istanbul~University, Science Faculty, Department of Astronomy and Space Sciences, Beyazıt, \.Istanbul 34119 Turkey}

\author[0000-0002-5274-6790]{Ersin G\"o\u{g}\"u\c{s}}
\affiliation{Sabanc\i~University, Faculty of Engineering and Natural Sciences, \.Istanbul 34956 Turkey}

\author[0000-0002-4387-7684]{Mete Uzuner}
\affiliation{Sabanc\i~University, Faculty of Engineering and Natural Sciences, \.Istanbul 34956 Turkey}

\author[0000-0003-4452-479X]{Asl{\i}han M. \"Unsal}
\affiliation{Sabanc\i~University, Faculty of Engineering and Natural Sciences, \.Istanbul 34956 Turkey}

\begin{abstract}
The Gamma-ray Burst Monitor (GBM) on board \textit{Fermi} Gamma-ray Space Telescope has produced the largest database of all-sky observations in gamma rays with its continuous data with high time and energy resolutions. These data contain a wealth of unidentified transient events that did not trigger the detectors for various reasons.
We conducted extensive searches to identify such untriggered transient events observed with GBM in 11 years (July 2010 -- June 2021). In particular, we employed four different search modes with various energy ranges (mainly below 300\,keV) and time resolutions (from 8\,ms to 2\,s), utilizing three statistical methods (signal-to-noise ratio, Poisson, and Bayesian statistics), each with different effectiveness in identifying specific classes of transients.  Moreover, we developed algorithms for known-event flagging as well as unknown-event classification for our candidate events found in the searches.
In this paper, we present our search methodologies, event flagging and classification algorithms and the resulting comprehensive event catalog.  The catalog contains more than a million events in total, including known events such as gamma-ray bursts, soft-gamma repeater bursts, galactic X-ray source activities, terrestrial gamma flashes, and solar flares.
For each candidate event, the catalog presents the event time, detection significance, event duration, hardness ratios, known-event flagging results, and classification probabilities.
Our short-transient catalog significantly expands the currently-existing list of known events and complements the GBM trigger catalog.
The event database with filtering capabilities is also publicly available at \url{https://magnetars.sabanciuniv.edu/gbm}, which allows users to retrieve event information based on their input queries along with the event lightcurves.
\end{abstract}


\section{Introduction} \label{sec:intro}

Gamma-ray Burst Monitor (GBM) on board the \textit{Fermi} Gamma-Ray Space Telescope has been observing the entire sky in the gamma-ray energy range since August 2008 \citep{meegan2009}. GBM is designed to detect short transient events of astrophysical origins such as Gamma-Ray Bursts (GRBs) or bursts from neutron stars, by monitoring the sky with a large field of view. Continuous, high-time-resolution daily GBM data are publicly available for all dates starting on July 16, 2010 (with the full 24-hour coverage starting on November 26, 2012), providing a database with over 14 years worth of data.  It has observed more than 11,000 short transients, including $\sim$4000 GRBs to date, immensely increasing the size of the GRB database available to the scientific community since its launch.
These monitoring detectors accumulate a wealth of data every day that contain scientifically valuable astrophysical events.  However, many of these events in the data do not get noticed unless specifically searched for. Examples of such events are sub-threshold GRBs, X-ray bursts, and short bursts from magnetars -- strongly magnetized neutron stars, which are also termed Soft-Gamma Repeaters (SGRs).
These transient events may fall short of the GBM on-board triggering criteria due to their energetically soft or dim nature.  Additionally, if another burst occurs during the data readout time of a triggered event (within $\sim$600 s after the trigger), the latter event will not be able to trigger the detectors.
Fortunately, these events can be (and have been) identified via offline searches in the archival data of continuous gamma-ray observations.  

The GBM detectors are designed to detect primarily GRBs. However, some GRBs are observed with low counts or intrinsically soft, hence evading GBM triggers.
The GBM data have previously been utilized for uncovering these ``untriggered" GRBs.
Some GRBs are associated with bright afterglow, supernovae, or even gravitational waves. Their progenitor environments, be it a core collapse of a massive star or compact object mergers, should have a limited amount of available energy that is expected to be of the order of the rest mass of the object(s) involved. Therefore, detection of sufficiently bright counterparts (electromagnetic or gravitational waves) may mean that the corresponding prompt gamma-ray emission of such GRBs could be intrinsically less energetic than ordinary GRBs without bright counterparts and may not meet the triggering criteria of the detectors.  Moreover, since GRBs are emitted within jet structures, off-axis emission may be observed as weaker, softer GRBs, and high-redshift GRBs could naturally be fainter than nearby ones keeping their flux at the subthreshold level.
In fact, the short GRB associated with the gravitational wave event, GRB\,170817A, was an especially weak although it was triggered by GBM
\citep{goldstein2017}, and another short GRB potentially associated with another gravitational wave event, GRB\,150914, was also weak, only detected by an offline search \citep{Connaughton16}. Following this, a targeted offline search was conducted for short GRBs using continuous GBM data \citep{kocevski2018, goldstein2019}. Their search is targeted, meaning that it only searches for particular event times, which they applied to look for subthreshold GRBs coincident with gravitational wave events when detected. The search approach used in this “\textit{Targeted Search}” is based on the likelihood values of the expected count rates using a set of template spectra, which in turn can better constrain the event locations. This is different from the search approach that we used in this work presented here.

Apart from GRBs, untriggered SGR bursts have also been searched previously within the GBM data for sources such as SGR\,J0418+5729 \citep{vanderhorst2010}, SGR\,J1550–5418 \citep{kaneko2010, vanderhorst2012, collazzi2015}, and more recently SGR\,J1935+2154 \citep{lin2020a, lin2020b}. For some of these sources, the untriggered event searches revealed hundreds of bursts per source during their active bursting episodes. Note that these searches were performed only for a limited period of time, usually weeks to months around the corresponding triggered burst times. 
It is possible that weaker bursts from SGRs (including previously unidentified sources) could be found within the time periods that are never previously used for searches.
Also, the search methods and the data types employed in these previous studies varied depending on the conducted study. Untriggered events were traditionally searched based on the signal-to-noise ratio, which is the same approach as the GBM trigger algorithm. More recent studies used Bayesian block representations of lightcurves \citep{scargle1998, scargle2013} to identify untriggered events. Additionally, type-I X-ray bursts (i.e., thermonuclear bursts) from neutron stars were searched in the first three years of GBM data by
\citet{jenke2016}, which identified 752 bursts. These non-GRB transient events are softer than typical GRBs, and thus many of these bursts do not meet the triggering criteria of GBM in terms of energy ranges that are optimized to trigger on GRBs. Therefore, offline event search in softer energy ranges is necessary and vital to reveal softer (or weaker) SGR bursts and X-ray bursts.

These previous studies, however, focus only on specific types of events and times, use single identification method, or cover only the first several years of GBM operations.
An exhaustive blind search for untriggered transient phenomena within over ten years of GBM data would create a valuable database of untriggered events, covering various types of X-ray/gamma-ray transients, which would substantially expands the existing triggered event catalog. Such a search would have no predefined objectives and would employ multiple identification techniques with different ``triggering'' criteria in terms of energy and time scales in a systematic manner. Consequently, we conducted a comprehensive search for untriggered events in the extensive database of \fermi (11 years of data). Our thorough search, applying three independent statistical methods across different energy bands (within 10--1000\,keV) and time resolutions, provides a substantial complementary event sample for astrophysical events such as GRBs, magnetar bursts, X-ray bursts, and Terrestrial Gamma Flashes (TGFs).  Furthermore, we developed event-type classification and source identification algorithms, which were applied to the candidate events found in the search.  

In this paper, we present the search and event classification methods used in our comprehensive transient event search, along with a summary of the search results, which are accessible as a catalog.  We first describe the \fermi detectors and data types (\S{\ref{sec:gbm}}) as well as the on-board triggering criteria and the triggered event summary (\S{\ref{sec:trig}}). Then, we describe in detail the search and classification methodologies (\S{\ref{sec:method}}) and present the overall statistics of our results in \S{\ref{sec:result}}.
Our search was performed on the whole 11-year GBM high time resolution database (July 16, 2010 -- June 30, 2021). 
To our knowledge, no such comprehensive search for transient events has been performed previously using the GBM data. Therefore, our investigations presented here remarkably extend the samples of all types of gamma-ray transient events, which we share publicly through a web portal (\url{https://magnetars.sabanciuniv.edu/gbm}).  


\section{GBM Instrument and Data Types} \label{sec:gbm}

The \textit{Fermi} Gamma-ray Space Telescope, launched on June 11, 2008 into a low Earth orbit, is a space observatory designed to study gamma-ray sources.  It carries two science instruments: the Large Area Telescope (LAT; \citealt{atwood2009}) which observes gamma-rays from $\sim$20 MeV to $\sim$300 GeV and GBM \citep{meegan2009}, which observes at lower energies in the range from $\sim$8 keV to $\sim$40 MeV, and thus provides a broadband (over seven decades in energy) spectral information of a wide variety of astrophysical sources.
The primary objective of GBM is to detect and locate GRBs occurring at random locations in the sky with its wide field of view (8\,sr un-occulted by Earth). The GBM consists of twelve thallium-activated sodium iodide (NaI) scintillation detectors in the energy range from $\sim$8 keV to $\sim$2 MeV and two bismuth germanate (BGO) scintillation detectors in the energy range from $\sim$200 keV to $\sim$40 MeV. 
The 12 NaI detectors (numbered from 0 to 11) are positioned as four groups of three detectors each to acquire the whole sky field of view, and the BGO detectors are placed at the opposite sides of the spacecraft so that at least one of them can detect an event if it occurs above the horizon.  When a burst is detected (i.e., ``triggers" the detectors), GBM roughly computes the location of the triggered event by using the relative count rates of these detectors.  This allows the immediate reorientation of spacecraft to enable LAT to observe possible delayed high-energy emission of bright bursts. 

There are three types of continuously-accumulated GBM data: CTIME data with high time resolution (256 ms) and low energy resolution (8 energy channels), CSPEC data with low time resolution (4096 ms) and high energy resolution (128 channels), and CTTE (Continuous Time-Tagged Event) data with a resolution of 2 $\mu$s and 128 energy channels, which became available on July 16, 2010 with partial coverages and after November 26, 2012 with a full, continuous coverage of all orbits \citep{vonkienlin2020}.
The GBM instrument is triggered if the Flight Software (FSW) detects an increase in the count rates of two or more NaI detectors exceeding a set of predefined threshold values. This threshold is an adjustable value determined in units of the standard deviation ($\sigma$) of the background rate, which is an average of the accumulated count rates over the previous 17 seconds, excluding four seconds close to the trigger time \citep{meegan2009}. When a trigger occurs, the FSW switches the data output of CTIME and CSPEC data to higher time resolutions of 64 ms and 1024 ms, respectively,
from the above-mentioned  values for (nominally) 600 seconds after trigger, and also generates a trigger TTE data file (nominally 330 seconds long), starting 30 seconds before the trigger.
Detailed information of all GBM triggers are accessible through the \fermi Trigger Catalog (FERMIGTRIG)\footnote{\label{trigcat}\url{https://heasarc.gsfc.nasa.gov/w3browse/fermi/fermigtrig.html}}, which includes trigger name, trigger time, triggered detectors, trigger algorithm number, and localization and classification of the trigger with the probability for the classification. The FERMIGTRIG is a dynamic catalog and automatically updated as new triggers are identified.
Finally, we note that the GBM detectors are routinely shut down during the spacecraft’s transit through the South Atlantic Anomaly (SAA), a region where the inner Van Allen radiation belt comes closer to the Earth surface and exposes the spacecraft to intense radiation and high particle activity; therefore, the count rates of the detectors goes to zero when the spacecraft passes through this region.


\section{\fermi Triggered Events}\label{sec:trig}

Both astrophysical and non-astrophysical events can trigger \ferminosp.
Besides GRBs, events that trigger \fermi can be classified into the following categories: Soft Gamma Repeaters (SGR), Solar Flares (SFLARE), Terrestrial Gamma-ray Flashes (TGF), Transients (TRANSNT), Local Particle events (LOCLPAR), Distant Particle events (DISTPAR), and Uncertain classification (UNCERT)$^{\ref{trigcat}}$.  A brief description and the time and energy profiles of each event are given below.

\textbf{GRBs} are intense and non-repeating gamma-ray transient events associated with the formation of black holes. 
GRBs come in a wide variety of shapes and duration, making it difficult to classify them morphologically.  However,
they are usually further classified into two classes, short GRBs  (SGRBs) and long GRBs (LGRBs) based on their duration \citep{kouveliotou1993}, and the two classes are thought to differ by the nature of their progenitors. SGRBs have observed duration of less than $\sim$2 seconds and likely originate from the mergers of two compact objects such as black hole-neutron star binaries and double neutron star binaries \citep{eichler1989,narayan1992,mochkovitch1993} in early type or star-forming galaxies \citep{bloom2006,fong2013}. 
LGRBs, however, last longer than $\sim$2 seconds (up to $\sim$1000\,s) and are likely associated with massive stars' cores collapsing into black holes \citep{woosley1993} or neutron stars \citep{thompson2004} with Type Ic core-collapse supernovae \citep{woosley2006,macfadyen2001} in late-type star-forming galaxies \citep{bloom2002,wainwright2007}.
SGRBs tend to be spectrally harder than LGRBs although there is no clear correlation. The GRB spectra are typically characterized by smoothly-broken power laws (of average indices $\sim-1.1$ and $\sim-2.2$) with an average break energy of $\sim100-200$\,keV \citep{kaneko2006,poolakkil2021}.

\textbf{SGRs}, also known as magnetars, are young,  slowly spinning (\textit{P} = 2$-$12\,s) neutron stars associated with extreme magnetic fields ($B >$ 10$^{14}$ G) \citep{kouveliotou1998,gogus2010}. Magnetars’ spatial distribution in the sky is strictly confined to the Galactic plane, which means that they are galactic sources. Spontaneous decay of the enormous magnetic field is thought to provide an energy source for short bursts as well as persistent emission in the X-ray and gamma-ray bands and long outbursts \citep{duncan1992,thompson1995,thompson1996}. The most common type of observed magnetar events is short bursts with duration ranging from a few milliseconds to a few seconds \citep{gogus2001} with the energy release of over $\sim$ 10$^{38}$ erg. In addition, short bursts tend to be single-peaked and decay slowly compared to the rise \citep{gogus2001}, although some short bursts with peculiar morphologies have been observed \citep[e.g.,][]{Israel2008, kaneko2021}. There are two more types of SGR bursts: giant flares and intermediate bursts. The giant flares consist of an initial spike releasing $>$10$^{44}$ erg of energy followed by a pulsating tail that decays in a long period. They are spectrally harder, peaking in the soft gamma-ray band and extending at least to MeV energies \citep{kaspi2017}. The intermediate events are between the short bursts and giant flares in terms of both duration and energetics, lasting a few seconds to a few tens of seconds, releasing up to $\sim$ 10$^{42}$ erg of energy \citep{kozlova2016}.  As the name suggests, SGR bursts are observed in the soft gamma-ray band below $\lesssim$\,200\,keV, with the $\nu F_{\nu}$ spectra peaking around $\sim 50-100$\,keV.

\begin{figure}[ht!]
    \centering
    \includegraphics[width=0.9\textwidth]{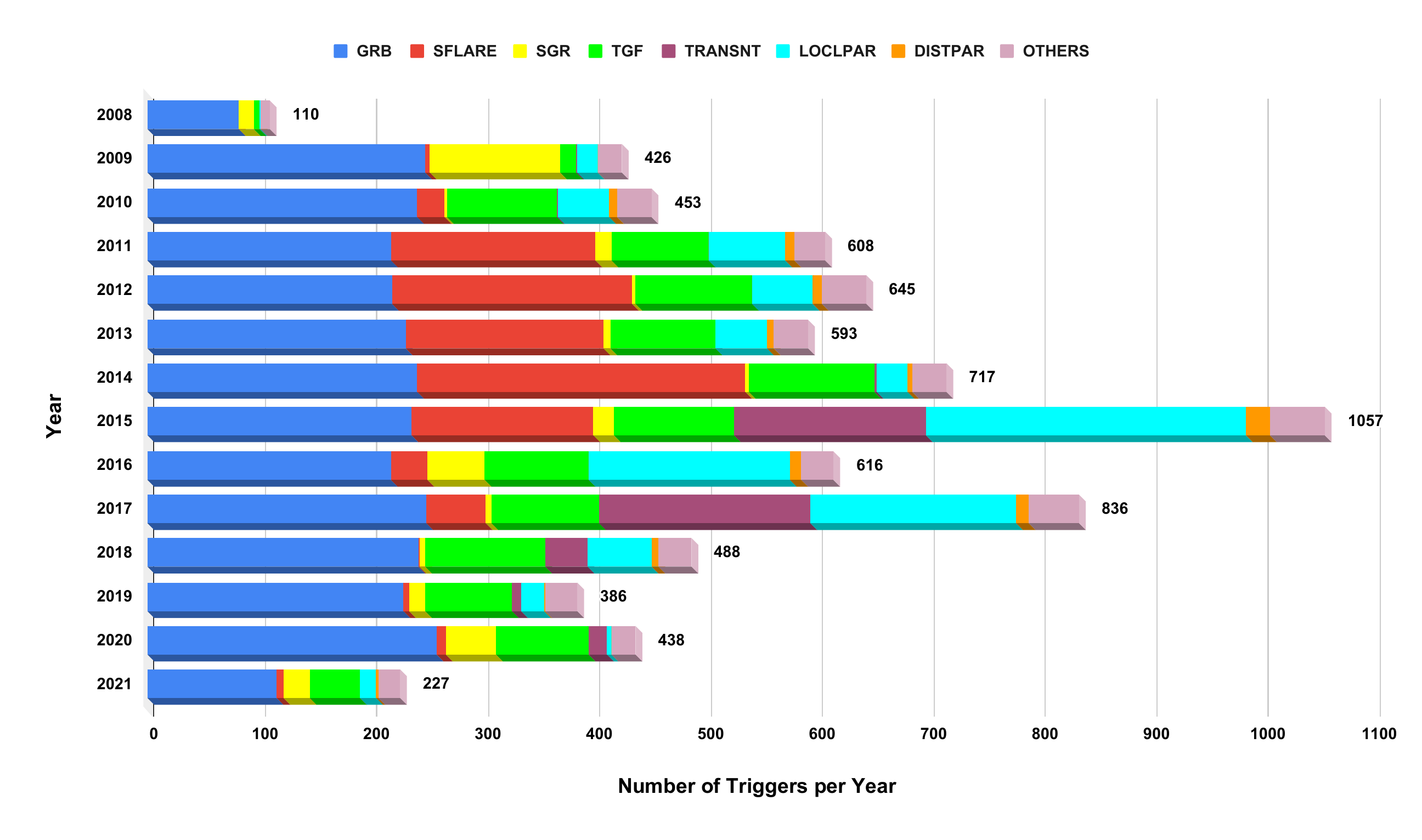}
    \caption{Trigger history of \fermi up to June 30, 2021 (i.e., our search period). Note that the number of triggers in 2008 and 2021 is not for a complete year.  The ``OTHERS" category includes UNCERT and Galactic Binary (GALBIN, only three events classified as such).}
    \label{fig:gbmhist}
\end{figure}

\textbf{SFLAREs}  can be quite energetic (in excess of 10$^{32}$ erg of energy), releasing energy across the entire electromagnetic spectrum from radio waves to high-energy gamma-rays.  Their duration can vary from seconds to hours \citep{fletcher2011}.  The gamma-ray spectra of SFLAREs generally span a wide energy range, and they have a variety of time profiles.
The SFLARE occurrence frequency depends on the 11-year solar cycle, and most of the SFLAREs \fermi observed so far occurred between 2010 and 2017, around the solar maximum of the 24th solar cycle (see \autoref{fig:gbmhist}).
SFLAREs are observed in all detectors facing the Sun at the time of the events.

\textbf{TGFs} are ultra-intense sub-millisecond gamma-ray pulses of atmospheric origin \citep{briggs2013} discovered by the Burst and Transient Source Experiment on the \textit{Compton} Gamma-Ray Observatory \citep{fishman1994}. 
The energy spectra of TGFs are hard, reaching up to several tens of MeV \citep{briggs2010}. Some TGFs are observed to have tails longer than a millisecond in duration \citep{roberts2018}.
Due to their hard nature, many TGFs detected with \fermi have been triggered using BGOs, which are sensitive in the MeV range.

\textbf{TRANSNT} events in \fermi Trigger Catalog are defined as non-SGR bursting activity of Galactic sources \citep{vonkienlin2020}. These sources can also go through outburst periods observed in soft gamma-rays; some of them such as pulsars, emit distinctive periodic bursts observed in energy below $\sim$50\,keV during the outburst periods.

Finally, \textbf{LOCLPAR} and \textbf{DISTPAR} are particle events due to charged particles of magnetospheric origin or occasionally cosmic ray showers taking place near or within the spacecraft. The charged particles trapped by the Earth’s magnetic field in Van Allen Belts can trigger the detectors during the spacecraft’s transit through the entry or exit regions of SAA or at high geomagnetic latitude \citep{vonkienlin2020}.  
Events that occur close to the spacecraft are classified as LOCLPAR. Such particle events are simultaneously detected by most of the GBM detectors and hence the light curves of 12 NaI detectors are quite similar in the 50--300\,keV range \citep{jenke2016}. In contrast, DISTPAR events occur near the Earth’s horizon and hence can be seen by only a few detectors unlike LOCLPAR events. Particle events in both types are mostly triggered with longer time scale (2.048 or 4.096\,s) algorithms in soft energy bands.
Note that these particle events are not of our scientific interests; nonetheless, they often trigger the GBM detectors and must be included in the trigger event category.

The on-board triggering algorithm of the GBM detectors is based on Signal-to-Noise Ratio: SNR \(= (R_{T}-R_{B})/\sigma\), where $R_{T}$ is the total count rate, $R_{B}$ is the background count rate, and \(\sigma = \sqrt{N_{B}}/\Delta t\), where $N_{B}$ is the total background counts in a data bin and $\Delta t$ is the duration of the data bin. 
The GBM triggering threshold varies from 4.5$\sigma$ to 8$\sigma$ in four possible energy bands, 25 – 50, 50 – 300, $>$100, or $>$ 300 keV, constituting a total of 119 preset triggering algorithms concurrently in use, including four algorithms based on the BGO detector count rates \citep{vonkienlin2020}. 
To identify a trigger, two or more detectors need to simultaneously record count rates exceeding the threshold so as to exclude possible triggers due to non-astrophysical sources.
As stated in the previous section (\S\ref{sec:gbm}) when there is a trigger, higher resolution trigger-mode data are accumulated for $\sim$300–600 seconds. During the trigger mode data acquisition, no further trigger can be recorded.
Within the 11-year time period covered in our search (July 16, 2010 – June 30, 2021), GBM triggered a total of 6811 events, of which 38\% are officially classified as GRBs, 2.8\% SGRs, 15.5\% TGFs, 16.8\% SFLAREs, 6.3\% TRANSNT, and 15.2\% LOCLPAR/DISTPAR via its on board classification algorithm or by the GBM team (see \autoref{fig:gbmhist} for the yearly breakdown of the triggered events).

\section{Search and Classification Methodology} \label{sec:method}

Our comprehensive event search process consists of two major steps; \textbf{(\textit{i})} searching the entire \textit{Fermi} GBM continuous high-time-resolution data to identify all short transient events using various statistics with multiple time resolutions and energy ranges (\S\ref{subsec:search} \& \ref{subsection:coverage}), and \textbf{(\textit{ii})} classifying the events detected by our search algorithm (\S\ref{subsec:classify}). We describe the methodologies that we employed for each step in this section.

\subsection{Transient Event Search and Candidate Identification}\label{subsec:search}
 
We used four different sets of time resolutions and energy ranges for identifying events within the continuous lightcurves of the 12 NaI detectors.
\begin{table}[!h]
\centering
\caption{Time resolution and energy range selections for each of the four search modes, applicable to all three search methods (SNR, Poisson, and Bayesian)}
\begin{tabular}{| c | c | c |}
\hline
\textbf{Mode} & \textbf{Time resolution} & \textbf{Energy range} \\
 &  \textbf{(ms)} &  \textbf{(keV)} \\
\hline\hline
\textbf{1} &  8 & 10 – 100 \\
\hline
\textbf{2} &  512 & 10 – 300 \\
\hline
\textbf{3} &  16 & 50 – 1000 \\
\hline
\textbf{4}\textbf{\textsuperscript{†}} &  2048 & 10 – 25 \\
\hline
\end{tabular}

\label{tab:modes}
\tablecomments{\textsuperscript{\textbf{†}}
Also used as a computational-check mode}
\end{table}
These four search ``modes" are listed in \autoref{tab:modes}, which we determined as follows:
In the blind transient event search using the GBM data, we expect to find several distinct classes of events, namely GRBs, SGR (or magnetar) bursts, TGFs, X-ray transients such as X-ray Bursts (XRBs), and Solar Flares (SFLAREs), as described in \S\ref{sec:trig}. Scientifically, we are interested in identifying energetically softer events, such as SGRs and XRBs, which are less studied with the GBM data compared to GRBs and other gamma-ray events. Based on these, we first studied the profiles of all types of GBM triggered events and the corresponding triggering algorithms.  
GBM was triggered more than 7400 times since the beginning of the mission (up to the end of 2020), the majority ($\sim$70\%) of which are classified as either GRBs, TGFs, or SFLAREs (see \autoref{fig:gbmhist} and {\it Fermi} GBM Trigger Catalog\footnote{ \url{https://heasarc.gsfc.nasa.gov/w3browse/fermi/fermigtrig.html}}). The triggers identified as SGRs comprise $\sim$4\%. We studied the triggering time scales and energy channels for each of these classification types and found the common triggering scales shown in \autoref{tab:trig_criteria}. Accordingly, we selected four sets of the most optimum time resolution and energy range that cover lower energy ranges than the actual triggering scales, aimed at finding a wide variety of event types 
(\autoref{tab:modes}).
The resulting definitions of the four ``Modes" complement the on-board triggering algorithms of GBM, so that our search results will supplement and add to the existing short-transient database.
\begin{table}[h]
\centering
\caption{The GBM triggering time and energy scales most commonly applicable for each of the event types for GBM-triggered events}
\label{tab:trig_criteria}
\begin{tabular}{|c|c|c|}
\hline
\multirow{2}{*}{\textbf{Triggered Event Type}} & \multicolumn{2}{c|}{\textbf{Common Triggering Scales}} \\
\cline{2-3}
 & \textbf{Time Resolution (ms)} & \textbf{Energy Range (keV)} \\
\hline\hline
GRBs & 64 – 4096 & 50 – 300 \\
\hline
SFLAREs & 64 & 25 – 300 \\
\hline
TGFs & 16 & 300 – 2000 \\
\hline
SGRs & 16 – 64 & 25 – 300 \\
\hline
Particle Events & 2048 – 4096 & 50 – 300 \\
\hline
Transients & 64 – 4096 & 50 – 300 \\
\hline

\end{tabular}

\end{table}

For candidate event identifications in each of the four modes, we employed three independent algorithms: 1.~signal-to-noise-ratio (SNR) based, 2.~Poisson statistics based, and 3.~Bayesian statistics based methods. Each method has its own unique strengths and weaknesses; thus, applying multiple search methods would result in identifying the maximum number of untriggered events overall.  
However, note that in this work, we present the results of each search method separately without incorporating a combined across-method statistics.
Below, we describe the three event identification methods.


\subsubsection{Signal-to-Noise Ratio (SNR) Method}

\begin{figure}[htbp]
    \centering
    \includegraphics[width=0.49 \textwidth,  trim=10 275 10 8, clip]{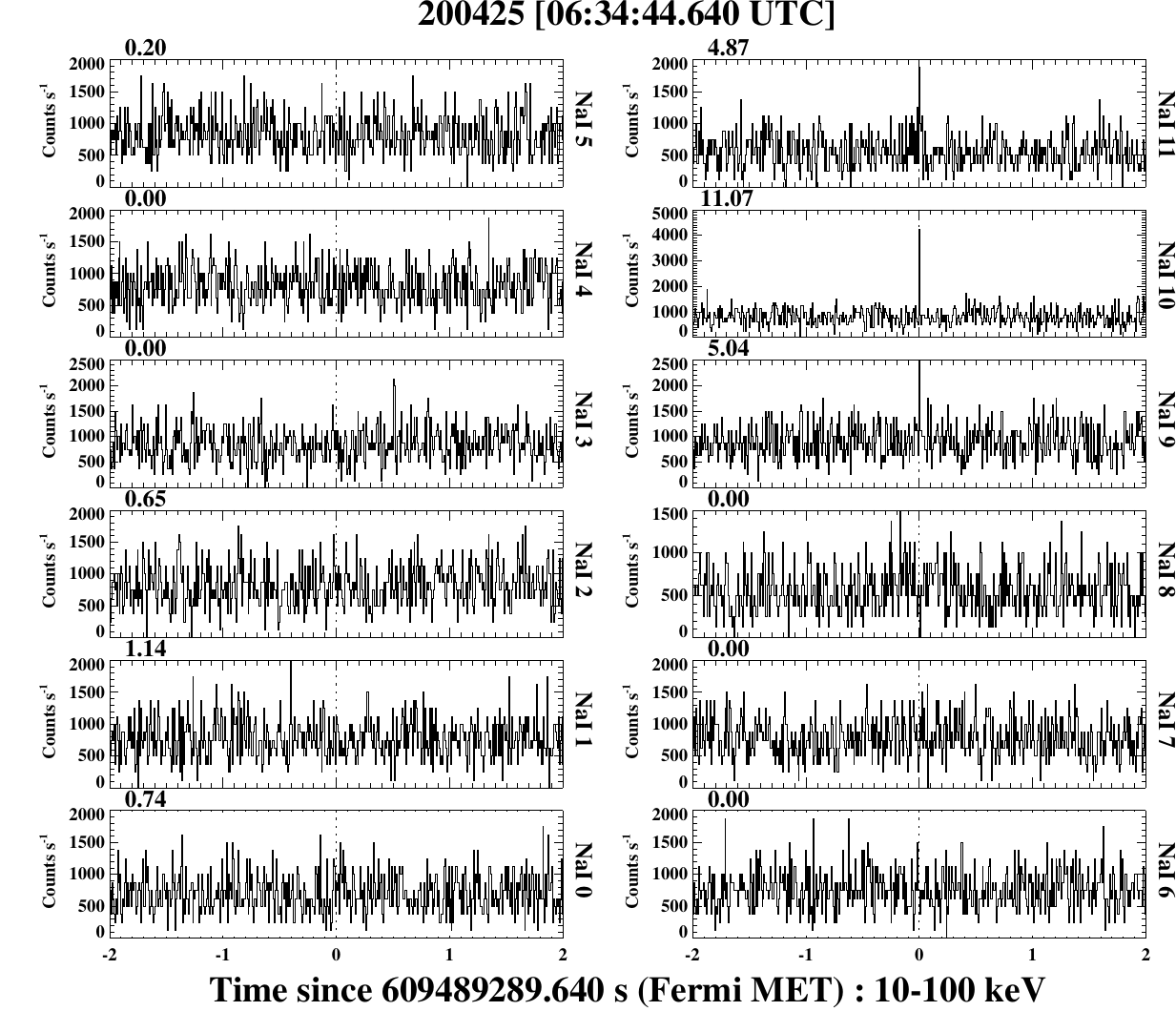}
    \includegraphics[width=0.49 \textwidth,  trim=10 275 8 8, clip]{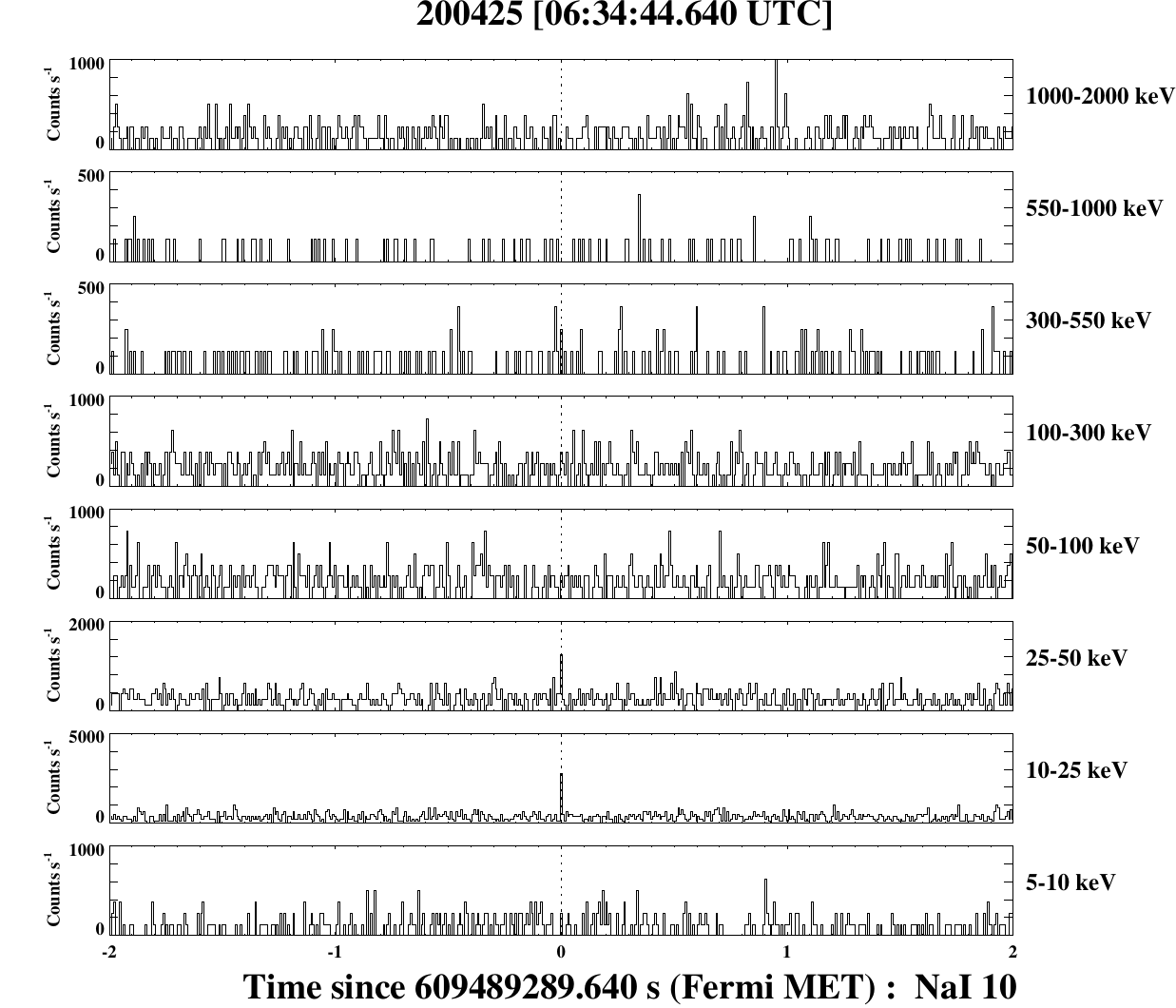}
    \caption{A burst flagged in the SNR search (likely from a magnetar, Swift\,J1818.0--1607).  [Left panel] The event is detected in three detectors (NaI\,9, 10, 11) in Mode-1 energy range (10--100 keV), with SNR = 5.0$\sigma$, 11.1$\sigma$ and 4.9$\sigma$, respectively.  The numbers at the top left of each panel are the significance in $\sigma$. [Right panel] The corresponding energy-dependent ligthcurves for the brightest detector, NaI\,10.}
    \label{fig:snr-example}
\end{figure}

Our SNR method is essentially the same approach as the GBM trigger algorithm (described in the previous section; SNR $= (R_{T}-R_{B})/\sigma$, where $\sigma = \sqrt{N_{B}}/\Delta t$) but with the SNR threshold fixed to the lowest value (4.5$\sigma$) to increase the sensitivity.  Note that a threshold lower than 4.5$\sigma$ would drastically increase the false detection rate. The SNR search algorithm looks at each time bin of the continuous data one by one for the 12 NaI detectors individually. It models the background count rates by linear fitting to a 10-second stretch of data before and after the time bin being evaluated. We left out a time window of 3 seconds immediately before and after the bin under evaluation so that it would not contribute to the background.  If data were incomplete $\pm 13$\,s of the time bin under evaluation, we used only available data either before or after for the background rate estimation.  Once the background rate is estimated, the algorithm then calculates corresponding SNR and flags the time bins with the SNR above the defined threshold (4.5$\sigma$) simultaneously in two or more detectors. If consecutive multiple bins are flagged, they are collectively identified as one transient event.  An example short transient event identified with the SNR method (in Mode 1) is shown in \autoref{fig:snr-example}.


\subsubsection{Poisson Statistics Method}

In the Poisson statistics search method, the event identification is based on the Poisson probability instead of SNR values.  The Poisson probability is calculated as follows: 
For each data segment with a certain number of time bins, the number of counts in the i$^{th}$ time bin will be compared to a local mean ($\lambda_i$). The local mean will be calculated from a 10-second stretch of data before and after the time bin being evaluated. As in the SNR method described above, we left out a time window of 3 seconds immediately before and after the bin under evaluation so that it would not contribute to the local mean. For the comparison, we determined the chance probability ($P_i$) of the number of counts in a time bin ($n_i$) being greater than the $\lambda_i$: 
\[P_i = \frac{\lambda_i^{n_i} e^{-\lambda_i}} {n_i!} .\]

\begin{figure}[htbp]
    \centering
    \includegraphics[width=0.6\textwidth]{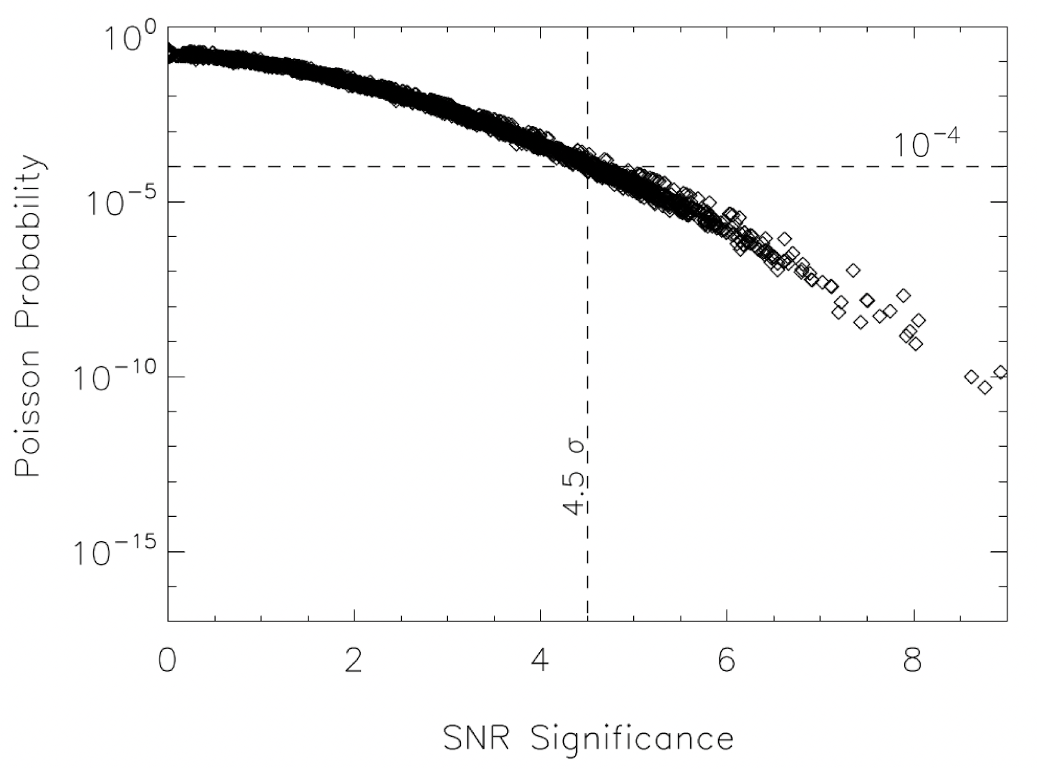}
       \caption{Calculated Poisson probabilities of all 12 detectors as a function of SNR for 220 events detected by the search with SNR method. The searched data were binned to 8-ms in the energy range of 10–100 keV (i.e., Mode 1).  $P \approx 10^{-4}$ corresponds to SNR = 4.5$\sigma$.}
    \label{fig:ps_sn}
\end{figure}

\begin{figure}[htbp]
    \centering
    \includegraphics[width=0.49 \textwidth,  trim=10 275 10 8, clip]{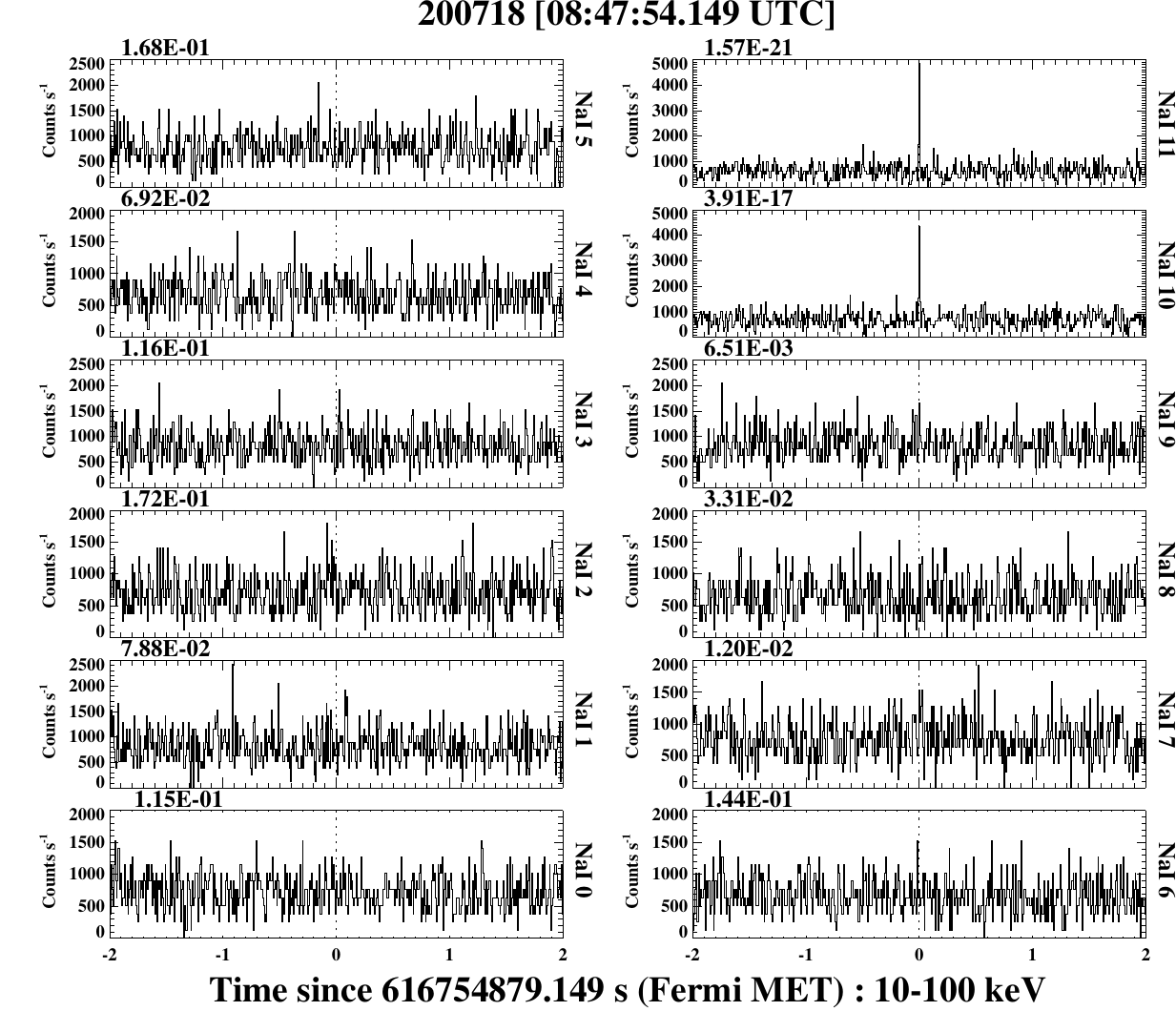}
    \includegraphics[width=0.49 \textwidth,  trim=10 275 8 8, clip]{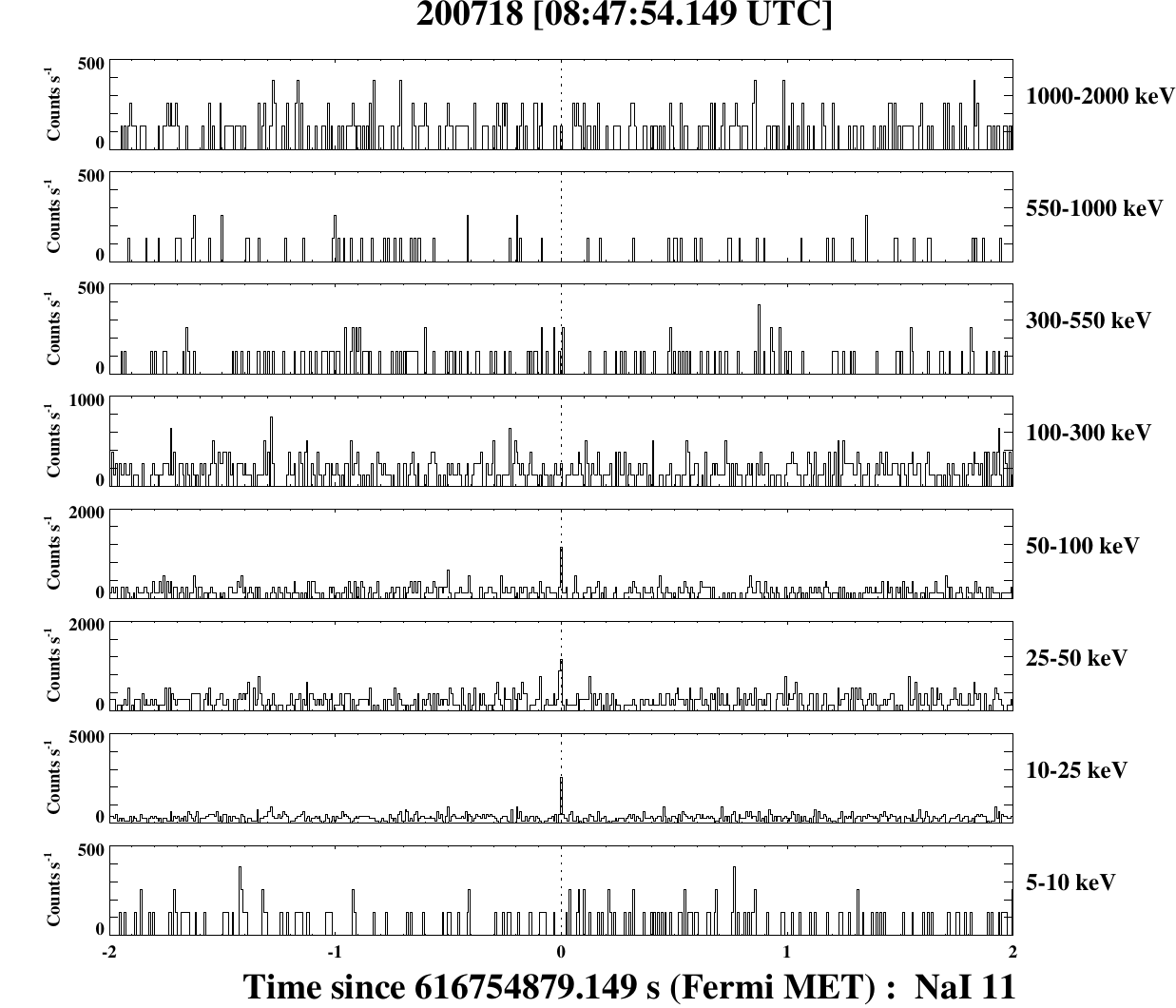}
    \caption{A burst flagged in the Poisson search (likely from a magnetar, PSR J1846.4--0258).  [Left panel] The event is detected in two detectors (NaI 10 \& 11) in the Mode-1 energy range (10--100 keV), with $P$ = 3.9E--17 and 1.6E--21, respectively. The numbers at the top left of each panel are the Poisson probability. [Right panel] The corresponding energy-dependent ligthcurves for the brightest detector, NaI\,11. }
    \label{fig:poi-example}
\end{figure}

When we have $P \leq 10^{-4}$ in two or more detectors the time bin is flagged as a candidate event. As in the case of the SNR method, consecutive bins qualifying the probability criteria were collectively identified as one transient event. The Poisson probability threshold of $10^{-4}$ was chosen since it roughly corresponds to a SNR of 4.5$\sigma$, which is the lowest threshold we use for the SNR search method as well as in the GBM on-board trigger algorithm. 
To determine the threshold, we used our initial SNR (Mode~1) search results of 26 days for which 220 events were found.  We manually calculated $P_i$ for the initial time bins of these events for all 12 NaI detectors' data (\autoref{fig:ps_sn}), and found the corresponding $P$ value for 4.5$\sigma$.
Above this threshold, we found that false positive detection becomes much more frequent. We note that this is comparable to the threshold $P$ value used for an offline search for TGFs performed by the GBM team ($P \leq 10^{-3}$ per detector, \citealt{briggs2013}).
An example event identified only with the Poisson method search is shown in \autoref{fig:poi-example}.


\subsubsection{Bayesian Statistics (Bayesian Block) Method} \label{subsec:bayesian-method}

The last method that we employed in identifying transient events is the Bayesian statistics method, based on which the lightcurves are represented by series of Bayesian ``blocks".  It is a powerful method to detect structures in time series with high degree of variability \citep{scargle1998,scargle2013}, as it provides a simplified description of the overall morphology of (and around) a transient event. Additionally, it could yield a direct measurement of the event duration. This approach has been used to calculate the duration of GRBs detected with {\it Swift} Burst Alert Telescope \citep{lien2016}, as well as to search for the extended emission following short GRBs \citep{norris2010,norris2011,kaneko2015}, and magnetar bursts \citep{lin2013,kirmizibayrak2017,lin2020a,lin2020b}.

\begin{figure}[htbp]
    \centering
    \includegraphics[width=0.9 \textwidth,  trim=75 142 20 275, clip]{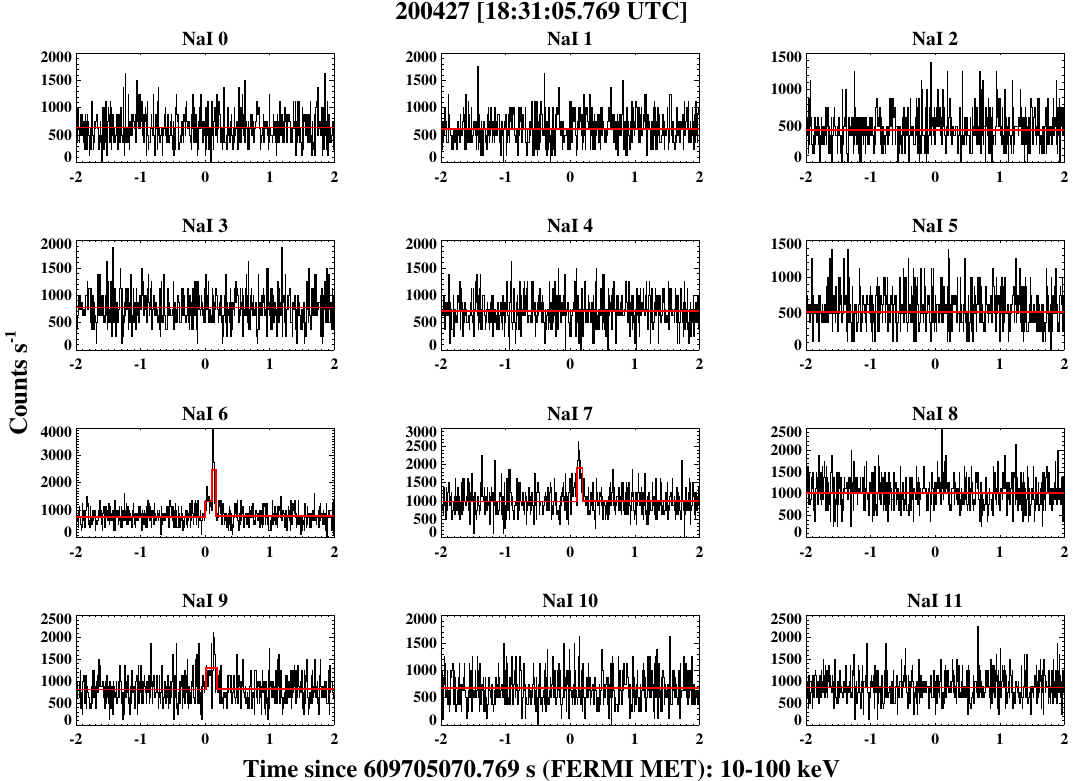}
    \caption{A burst flagged only with the Bayesian block search (likely from a magnetar, SGR J1935+2154).  The event is detected in three detectors (NaI 6, 7 \& 9) in the Mode-1 energy range (10--100 keV). The red lines show the Bayesian-block representation of each lightcurve, plotted over the 8-ms lightcurve.}
    \label{fig:bay-example}
\end{figure}

The Bayesian block method simply represents a time series with step functions of various widths (duration) and heights (intensity). The durations (or ``change points") of these steps are determined by maximizing the block likelihood function. For photon counting instruments such as \ferminosp, the probability of detecting an event follows Poisson statistics. The likelihood function ($L$) for block $k$ then is
\[\ln L(k) = N(k) \ln \lambda(k) - \lambda(k) T(k),\]
where $N(k)$ is the total number of photons in block $k$, $\lambda(k)$ is the expected count rate, and $T(k)$ is the duration of block $k$. A priori information is the improvement of the likelihood function by adding one more block, which depends on the number of photons, as well as the false positive change point probability of 0.05 per data set \citep{scargle2013}. This makes the Bayesian block method potentially sensitive to weak events, which may be smoothed out when the data are uniformly binned. Moreover, the Bayesian block method also allows us to systematically estimate the duration of transient events, although identifying the event block and the background blocks become increasingly challenging as the event flux decreases (see \S\ref{subsec:untrig_classification}).
After the continuous data were represented in Bayesian blocks, we applied the following event detection procedure:
\begin{enumerate}
    \item[i.] Since our aim is to find short transient events (such as SGR bursts), blocks $\leq$ 1 s (for Modes 1 and 3; $\leq$ 100 s for Mode 2 and $\leq$ 200 \,s for Mode 4) within each detector dataset are flagged as potential event blocks.  
    \item[ii.] The ``background" count rate is then determined by taking an average of the counts of longer blocks ($>$ 1 s or $>$ 100 and 200 s for Modes 2 and 4, respectively) immediately before and after the potential event blocks divided by the total duration of the blocks.
    \item[iii.] The potential event blocks that have higher count rates than the background count rate in two or more detectors simultaneously are identified as events. 
    \item[iv.] Consecutive event blocks are considered as one event. Additionally, two events that are separated by more than 0.1\,s are considered separate events.
\end{enumerate}
These criteria were inherited from our original Bayesian-block search algorithm designed specifically for finding sub-second short bursts (e.g., SGR bursts) in the data binned to 8\,ms or 16\,ms, so most of the events found with the Bayesian method in Mode 1 and 3 have short durations (79\% of the candidate events have duration $\leq$ 1 s in Mode 1, 89\% in Mode 3 (see \S\ref{sec:result} for more detailed statistics).
An example event identified only with the Bayesian method search (Mode 1) is shown in \autoref{fig:bay-example}.
 \\

For clarification, the event detection criteria for all three search methods (i.e., SNR, Poisson, and Bayesian block) described above are summarized in \autoref{search_criteria}.  The three search methods were applied to each of the four Modes (\autoref{tab:modes}); therefore, a total of 12 sets of search results for the 11-year worth of data were obtained.
\begin{table}[htbp]
\centering
\tabletypesize{\footnotesize}
\caption{Event detection criteria used for the searches with the three methods}
    \begin{tabular}{|l|l|}
        \hline
        \textbf{Method} & \textbf{Criteria for Event Detection} \\ 
                    & (required for two or more detectors simultaneously)\\
        \hline \hline
        Signal to Noise Ratio (SNR)&$>4.5\sigma$ above the background level\\ 
        Poisson statistics (Poisson)&Poisson probability $<10^{-4}$  \\ 
        Bayesian Blocks (Bayesian)&Any blocks above the estimated background level $^\dagger$ \\
        \hline
    \end{tabular}
    \tablecomments{\textbf{$^{\dagger}$}“Event” blocks are identified based on a pre-defined background duration, which is set here to $>$ 1\,s (or 100 and 200\,s for Modes 2 and 4, respectively) aiming to find short bursts. }
\label{search_criteria}
\end{table}
Finally, we note again that requiring two or more detectors to simultaneously meet the criteria for event identifications reduces the false detection due to random non-astrophysical signals.
It should be highlighted that our team has been successfully utilizing both the SNR method and the Bayesian method for searching untriggered events for several SGR events during their burst-active episodes \citep[e.g,][]{kaneko2010,vanderhorst2010,vanderhorst2012,lin2013,lin2020a,lin2020b,uzuner2023}.

\subsection{Data Coverage}\label{subsection:coverage}
GBM data are uploaded to the public data portal of NASA’s High Energy Astrophysics Science Archive Research Center (HEASARC) soon after it is downlinked from the spacecraft. We first downloaded all the daily CTTE data and the necessary calibration files to the local server for our comprehensive search and subsequent investigations. Since we aimed to uncover all detectable untriggered transient events in a semi-automated manner, it was crucial that the initial database does not include any missing or corrupted data; therefore, we first developed several data-management algorithms to check missing time intervals within the entire local GBM database for all 12 NaI detectors. The identified missing data were downloaded manually or noted if the original data on the NASA server were not available.  It is important to note that the continuous CTTE data became available on July 16, 2010, thus our search period started from this date.
Therefore, the transient event searches were performed using the CTTE data of all 12 detectors for the 11-year time period between July 16, 2010 and June 30, 2021, except for 13 days (March 16--28, 2018)\footnote{\url{https://fermi.gsfc.nasa.gov/ssc/data/access/gbm/gbm_data_gaps.html}; note that the data are partially available for March 16 and 28, right before and right after the anomaly period and thus, included in our search} during which the data for all NaI detectors are not available due to a \textit{Fermi} spacecraft operational anomaly \citep{vonkienlin2020}.

\subsection{Candidate Event Classification}\label{subsec:classify}

Subsequent to the event candidate identification by the search, the event candidates need to be evaluated for their origins and characteristics to classify their natures. Many of them are likely to be either known local events of non-astrophysical origins or artificial count rate increase due to SAA region entries and exits.  Some others may be astrophysical in nature; however, they include multiple detections of the same long-lasting events such as solar flares or long-duration GRBs, GBM-triggered (and thus already classified) events, or other cataloged events existing in literature or otherwise recorded.

Therefore, we developed algorithms to filter out known non-astrophysical events and flag known astrophysical events (including GBM triggered events, documented solar flares, active X-ray sources, etc.), as well as to classify events that are not previously reported. 
Below, we describe our approaches to known-event filtering algorithms (\S\ref{subsec:filtering}), and the unknown-event classification procedure (\S\ref{subsec:untrig_classification}).

\subsubsection{Known Event Filtering \& Flagging}\label{subsec:filtering}

\textbf{SAA Filtering:}
The \emph{Fermi} satellite is in a low-Earth orbit and makes $\sim$15 orbits around the Earth every day. Depending on the orbital path, it passes over the SAA, within which an excessive amount of charged particles are found. In order to avoid damage to the detectors, all detectors are turned off during the SAA passages. However, right before and after the SAA, the count rates could artificially shoot up, which resembles a short transient event. Therefore, we first identified all the candidate events that occurred $\pm$60 s of SAA passages and filter them out as potential false detections. About 11\% of candidates found in all 12 searches fall into this category.  We separately compiled the list of these filtered events with their event times, triggered detectors and the detection significances, and provide it in our web portal (also in \citealt{zenodo-catalog}), since they are excluded from our candidate event catalog.
Note that some of the excluded events could be known events, in which case we include the known event flag (see below for the descriptions) in the list.

\begin{figure}
    \centering
    \includegraphics[width=0.9 \textwidth,  trim=60 14 85 48, clip]{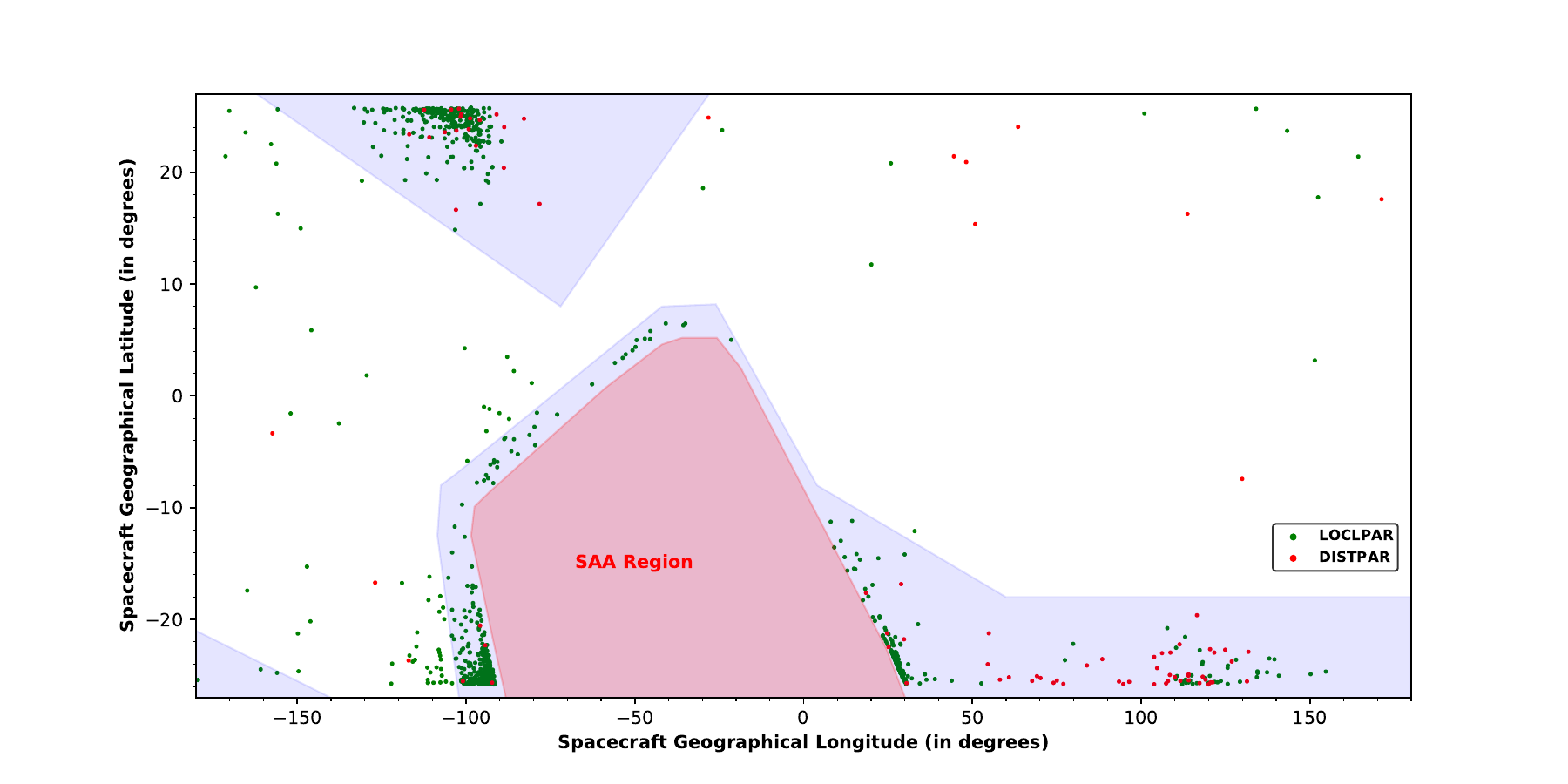}
    \caption{\emph{Fermi} spacecraft geographic coordinates at the times of GBM-triggered local particle events (green dots) and distant particle events (red dots).  The SAA region defined by \emph{Fermi} team is shown in red.  The blue shades show the regions with $L \geq$ 1.3 and the SAA entrance region, which we used to flag our candidates as particle events.}
    \label{fig:particle-location}
\end{figure}

\textbf{Particle Events:}
We also flagged all potential particle events within our event candidates using the spacecraft's geographical latitude and longitude at the time of each candidate event. The particle events mostly originate from the charged particles trapped by the Earth’s magnetic field, and hence occur at the edge of the SAA region or higher McIlwain $L$ coordinates ($L \geq$ 1.3) as shown in \autoref{fig:particle-location}. Therefore, we flagged (with flag = 1) the candidate events that occurred when the spacecraft was located at $L \geq 1.3$ or at the SAA entrance region as potential particle events.
Here, we only flagged the particle event candidates but did not exclude them from our candidate event catalog because these criteria are not decisive but indicates elevated probability of the event being a particle event.

\vspace{12pt}
Published short transient events, such as GBM-triggered events and cataloged solar flares, as well as activities from active X-ray/gamma-ray sources with known active periods and locations are also included in our candidate events and can also be flagged.  To this end, we first compiled a database of known transients during the last 11 years (from July 2010 to June 2021), which could be used for identifying previously known events among the candidate events found in our search. The database includes the event times, event classifications, duration, observed detectors, etc.~of all GBM triggers in the GBM Trigger Catalog as well as cataloged solar flares and active X-ray sources that are publicly available.

\textbf{GBM Triggered Events:}
As stated in \S\ref{sec:trig}, GBM was triggered by 6811 events during our search period, of which $>$95\% have already been classified (i.e., event types and/or the sources were identified) on board and by the instrument team.
Since the triggered events' duration spans a wide range (from ms to minutes) and our search time scale and energy range may differ from the trigger time scale and energy range, identifying these triggered events among our candidate events is not straightforward.  Besides, to identify a triggered event, the candidate event must coincide both in time and space (i.e., the source direction) with the triggered event.  
All-sky monitors in gamma-ray band like GBM cannot provide high-accuracy source locations; however, the count rates of the 12 detectors strongly depend on the location of the source and can be utilized for estimating the approximate source location. 
Therefore, after testing various matching methods, we chose the following method to match our candidate events to the triggered events. For each of the candidate events, we first check whether the trigger time of our candidate event falls within the duration of any of the triggered events. If so, we compare our ``triggered" detectors (i.e., the detectors that satisfy our search criteria) with the detectors that actually triggered GBM, by comparing the significance (or count flux in case of the Bayesian method) of the detected event for each ``triggered" detector. We assign the grade of 3 if both time and two or more detectors with the highest significance match (i.e., the candidate event is very likely the triggered event), grade 2 if time and one detector match, and grade 1 if only time matches. In addition to the associated trigger grade for each event in our catalog, we provide the associated trigger number and class given in the \fermi trigger catalog in case of a match.

\textbf{Solar Flares (SFLAREs):}
To identify solar flares in our search, we utilized the GBM Solar Flare Catalog\footnote{\url{https://heasarc.gsfc.nasa.gov/W3Browse/all/fermigsol.html}}, which lists the start, end, and peak times, and four detectors with the highest counts (i.e., sunward-facing detectors at the time of the flare), of all solar flares observed with GBM (5110 SFLAREs including 1142 triggered flares during our search period). 
Similar to the GBM triggered event identification algorithm above, our SFLARE identification algorithm compares the times of our candidate events to the time intervals of the cataloged flares, and if they match it compares the detectors with the highest significance as determined in our search. Based on this, we provide a solar flare grade for each event in our database. We assign the grade of 3 if both time and three or more detectors match, grade of 2 if time and two detectors match, and 1 if only time, or time and one detector match.  Additionally, for grade-1 events, we calculated the 12 detectors' zenith-to-Sun angles at the time of the candidate events (since some SFLAREs last tens of minutes) and compared the smallest-angle detectors to our ``triggered" detectors, and grades were reassigned accordingly. 

\textbf{Known X-ray Source Activities:}
The outbursts of four bright X-ray transient sources: V404~CYG (in June 2015; \citealt{jenke2016_v404}) and Swift J0243.6+6124 (in November 2017; \citealt{wilsonhodge2018}), MAXI J1820+070 (in April 2018) and 1A 0535+262 (in November 2020; Colleen Wilson-Hodge, personal communication) triggered \fermi multiple times, which are also detected in our search. 
To identify the activities of X-ray sources in our search, we first check whether the times of our candidate events fall into the outburst periods of the sources. If so, we compare the highest-significance detectors with the detectors with the smallest zenith-to-source angles.
We assign the grade of 3 if both time and two or more detectors with the highest significance match, grade of 2 if time and one detector match, and grade of 1 if only time matches and the source is not under the Earth blockage.

\subsubsection{Unknown-Event Classification}\label{subsec:untrig_classification}

After filtering out and identifying (i.e., assigning grades of 3) the known events among our candidates, namely, the SAA events and grade-3 GBM triggerred events, SFLAREs, and X-ray transients, the remaining candidate events (including flagged particle events) were subjected to an event classification algorithm with a Bayesian probability approach, similar to the method used for the on-board classification of GBM triggers \citep{briggs2007}. 
The on-board classification is done by calculating the probabilities of the event being a GRB, SGR, SFLARE, TGF, X-ray transient, or a particle event based on the observed parameters such as the event's rough location, hardness ratio, count ratio, and the spacecraft's geomagnetic coordinates.  Building on this, our classification method seeks to utilize more refined event class properties.
Therefore, we first performed a detailed study of collective properties of the classified triggered events based on their classifications (see Appendix \ref{app1}), which we used as one of the a priori probability distributions of the Bayesian probability calculations applied to the remaining candidate events found in our search.  

The classification algorithm starts with a given set of prior probability for each event class, which we formed based on the number of triggered events in each class among all GBM events (up to the end of 2020) triggered with similar criteria to our four modes. The algorithm then uses the event duration and spectral hardness ratio (HR) distributions of triggered events (see \autoref{app1} for the details and examples) as inputs to calculate posterior probability of each class for the candidate event.  
Ideally, the event duration should be calculated using the Bayesian block method for each unknown event candidate; however, when this was attempted, many weak/short event candidates found in the SNR and Poisson searches are not found in newly-generated Bayesian block lightcurves and the duration could not be calculated. 
Besides, determining the background blocks and event blocks proves challenging without pre-knowledge of the event class or type.
Therefore, we defined the event duration for the SNR and Poisson events as the number of consecutive time bins multiplied by the time resolution of the search and for the Bayesian events, we kept the event block duration (as described in \S\ref{subsec:bayesian-method}) as the event duration. Note that this approach by default makes the minimum ``duration" of events the time resolution of the data for each search mode.

For HR, we calculate three HR values in 10$-$2000\,keV at the peak of the event using a pair of “pivot” energies ($E_{\rm piv}$ = 25, 50 or 75\,keV), which is the dividing energy for the hard and soft energy ranges. These pivot energies were chosen so as to distinguish similar event types (short GRB vs. SGR, long GRBs vs. SFLAREs, in particular) as described in \autoref{app1}.  Also, to ensure the reliability of the HR values, we kept only the well-constrained HR values with the associated errors within 60\% of the HR values.  We found that there are a number of events for which the HR could not be obtained or constrained; these include spectrally hard events (such as potential TGFs; see below for the TGF identification) or very soft events (such as X-ray bursts), and weak events with minimum number of photon counts.  In those cases, we flagged the event as ``NO EVALUATION".

Subsequently, we calculated the probability of each event being a certain event class, based on the Bayes approach, taking the duration and HR distributions of triggered events as a priori \citep{mitrofanov2004,briggs2007}.  
The posterior probability of the event $j$ being the class $i$ is:
\[P(C_i|D_j) = \frac{P(C_i)P(D_j|C_i)} {P(D_j)} \]
where \(P(C_i)\) is the a priori probability for $i$ = 1--4 for four types of event classes (i.e., Long GRB, SFLARE, Short GRB, and SGR), and \(P(D_j|C_i)\) is the probability for the parameter $D$ for event $j$ to be observed given that the event belongs to a class $C_i$. The total of all \(P(C_i|D_j)\) for four classes must be normalized to 1.
Here, we set a priori probabilities based on the statistics of the events that triggered GBM detectors with the trigger algorithms similar to our energy and time resolution for each mode (see \autoref{tab:apriori}). They were set such that $\displaystyle\sum_{i = 1}^4 P(C_i) = 1$, where $C_{1-4}$ corresponds to the four classes: LGRB, SFLARE, SGRB, and SGR.
\begin{table}[!htbp]
    \centering
    \caption{GBM trigger algorithms used to set a priori probability for the unknown-event classification}
    \label{tab:apriori}
    \begin{tabular}{|c|c|}
    \hline
    \textbf{Mode} & \textbf{GBM Trigger Algorithm}$^*$ \\ \hline \hline
       1  &  1--5, 22--26 \\
       2  &  6--17, 22--38 \\
       3  &  1--5, 43, 50, 116--118 \\
       4  &  12--17, 22--38 \\
\hline
    \end{tabular}
    \tablecomments{$^*$ The algorithm numbers are from Table\,2 in \citet{vonkienlin2020}}
\end{table}
Note that TGF class is not included here because TGFs are extremely short ($\sim$1$-$2\,ms) and very hard, with HR\(_{[E_{\rm piv} = 300 {\rm\,keV}]}\) = 1.15 on average, and could be flagged separately; solely for this purpose, we also calculated HR with $E_{\rm piv}$ = 300\,keV.
Additionally, we emphasize that our aim here is to provide the probability of the event being one of the four event types, which are short transient events of higher interest in the astronomy community.  As we described earlier, some of the events are flagged as potential particle or X-ray transient events; they may still be classified as one of the four event types with a high probability. Thus, the Bayesian probabilities presented in the catalog should be used with cautions, taking into account the other flags associated with the events of your interest.

To test our classification algorithm, we applied this approach and calculated the probability for nearly all triggered events till December 31, 2020, that are classified as SGR, GRB, SFLARE, or TGF.  A priori probability used was 0.25 for each of the four classes. We excluded TGFs that trigger only the BGO detectors (GBM algorithm number 119, \citealt{vonkienlin2020}) and combined SFLAREs that trigger the GBM multiple times due to their long durations.
We found that the false classification rate is 0.3\% on average using the Bayesian probability approach, as shown in \autoref{tab:false-rate}.  Note that here, we combined the short GRB (SGRB) and long GRB (LGRB) as just GRB due to the ambiguity in the definition of ``short" GRBs.  
\begin{table}[htbp!]
    \centering
    \caption{False classification rates of triggered, classified events using our Bayesian probability classification algorithm}
    \label{tab:false-rate}
    \begin{tabular}{|c|c|c|c|c|}
\hline 									
\textbf{Class}	&	\textbf{Total}	&	\textbf{TRUE}	&	\textbf{FALSE}	&	\textbf{\%}	\\ \hline\hline
SGR	&	291	&	289	&	2	&	0.7	\\
GRB	&	2169	&	2167	&	2	&	0.1	\\
SFLARE	&	744	&	739	&	5	&	0.7	\\ \hline
ALL	&	3204	&	3195	&	9	&	0.3	\\
\hline									    
    \end{tabular}
\end{table}
SGRBs tend to be spectrally harder than LGRBs but their HR values significantly overlap, both centered around 1 for $E_{\rm piv}\sim$  25$-$75\,keV \citep[see e.g.,][and also \autoref{fig:hr-comparison} in \autoref{app1}]{salmon22}.  Their duration distribution also exhibit overlap between $\sim$1$-$10\,s (see \autoref{fig:grb_dur_dist} in \autoref{app1}). Therefore, it is a challenge to classify GRBs with duration $\sim$1$-$10\,s with HR $\sim$1 as SGRBs or LGRBs.
Overall, for all event classes, the falsely-classified events are those in the tail distribution of either the HR or the duration.
For TGF identification, we required the event to have duration $<$ 16\,ms as well as a constrained HR with $E_{\rm piv} = 300 {\rm\,keV}$ but unconstrained lower pivot energy HRs, i.e., $E_{\rm piv}$ = 25 and 50 keV. With these criteria, the false TGF classification rate was 3\% (16 out of 528 triggered TGFs); our algorithm falsely classified all of these 16 TGFs as (short) GRBs.



\section{Search Results \& Transient Event Catalog} \label{sec:result}

We ran the transient event searches with three statistical methods (SNR, Poisson, Bayesian) and four modes (Modes 1--4; \autoref{tab:modes}) all in parallel on 11 years of GBM CTTE data of 12 NaI detectors; the search period spans a total of 3986 days. In \autoref{tab:search_summary}, we summarize the numbers of identified candidate events sorted by methods and modes, along with the number of events included in the catalog. As stated before, each method employs a set of criteria to identify events, after which we filtered out the events that are potentially false detections due to the spacecraft's entry to or exit from the SAA region.  Overall, about 11\% of the candidate events were classified as SAA and excluded, as seen in the table. The remaining events constitute our catalog of short transient events.  The average number of events found per day varies from a few to $\sim$75 depending on the method and mode. 
\begin{table}[!htbp]
    \centering
    \caption{Summary of the results of 12 searches (three methods and four modes) performed for the total search period of 3986 days  }
    \label{tab:search_summary}
    \begin{tabular}{|c|c|c|c|}
    \hline
	 & 	\multicolumn{3}{c|}{Number of Events}						 \\	
  \cline{2-4}
Mode	 & 	Identified	&	SAA Filtered\tablenotemark{$^a$} (\%)		&	in the Catalog	 \\	\hline \hline
\multicolumn{4}{|l|}{\textbf{Signal-to-Noise Ratio (SNR) Method} } \\		\hline											
1	 & 	135,841	&	7,937	 (5.8)	&	127,904	 \\	\hline
2	 & 	136,431	&	8,129	 (6)	&	128,302	 \\	\hline
3	 & 	67,180 &	3,267	 (4.9)	&	63,913	 \\	\hline
4	 & 	144,103	&	15,592	 (10.8)	&	128,511	 \\	\hline
\multicolumn{4}{|l|}{\textbf{Poisson Method}}	 \\	\hline											
1	 & 	133,532	&	10,087	 (7.6)	&	123,445	 \\	\hline
2	 & 	292,416	&	22,727	 (7.8)	&	269,689	 \\	\hline
3	 & 	52,867	&	4,157	 (7.9)	&	48,710	 \\	\hline
4	 & 	301,253	&	34,022	 (11.3)	&	267,231	 \\	\hline
\multicolumn{4}{|l|}{\textbf{Bayesian Method}}		 \\		\hline									
1	 & 94,437	&	9811 (10.4)	& 84,626	 \\	\hline
2	 &	52,448	&	32,704	 (62.4)	&	19,744	 \\	\hline
3	 &	27,928	&	3,915    (14)	&	24,013	 \\	\hline
4	 &	33,938	&	15,903   (46.9)	&	18,035	 \\	\hline
\end{tabular}
\hspace{0.5cm}\\[0.3cm]
\raggedright
\tablecomments{
The search range (in YYMMDD) is from 100716 to 210630; The data for 180317--180327 are not available due to \\textit{Fermi} spacecraft operational anomaly. 
There are additional six days (101028, 101029, 130823, 130824, 130825, 140914) for which there are no CTTE data available.}
\tablenotetext{a}{The events found within $\pm$60\,s of the spacecraft SAA passages were excluded in our event catalog but the event information of excluded events is provided as a supplementary material in the accompanying Zenodo repository \citep{zenodo-catalog} as well as at \url{https://magnetars.sabanciuniv.edu/gbm}.}
\end{table}
We note that the numbers presented in \autoref{tab:search_summary} includes duplicates, meaning that there are overlaps in the populations included in our transient event catalog. 

Overall, statistically-expected false positive rates are low.  
The false positive rate can be calculated for the SNR method as follows.  The chance probability of two detectors simultaneously having 4.5$\sigma$ excess is $10^{-11}$; therefore, for 11 years of data binned to 8, 512, 16, and 2048\,ms resolution (corresponding to each mode), the expected numbers of false detections are 0.5, 0.008, 0.25, and 0.002, respectively.  Given the number of events identified in our search in each mode, the false positive rate is less than $\sim 10^{-6}$ in each mode. 
Additionally, for the Poisson method, our detection criterion was $P \leq 10^{-4}$ per detector, so the chance probability of two-detector simultaneous detection is $\leq 10^{-8}$.  Similar to the SNR calculations above, the expected numbers of false detections in 11 years are 434, 13.5, 217, 1.7 for the four modes, corresponding to the false positive rates between $\sim$$10^{-3}-10^{-7}$.
Finally, for the Bayesian method, the false ``change-point" detection rate of 5\% per dataset was used in our search. Based on this, we calculated the false event detection rate as follows: the total number of change points generated by the Bayesian block algorithm for all 3986 days per detector averages to 1.56E6, 2.49E6, 7.09E5 and 1.94E6 for Mode 1, 2, 3 and 4, respectively.  Then, say for Mode 2, we detected 52,448 events, which is 2.1\% of the number of the total change points.  So, per detector, we estimate the false detection rate of Mode 2 to be $0.021 \times 0.05 = 0.001$, which corresponds to a chance probability of $10^{-6}$ for two-detector simultaneous detections.  For other modes, we found the probabilities of $9.2 \times 10^{-6}$, $3.9 \times 10^{-6}$, and $7.6 \times 10^{-7}$ for Mode 1, 3, and 4.
The search results (in \autoref{tab:search_summary} and in the Catalog) are presented separately for each search mode and method, so the false positive rates were also calculated separately.  Since we perform blind searches, we caution that these rates are merely the chance probabilities for detecting an event given the threshold, the number of detectors, and the number of time bins for which the search was performed.

\subsection{The Event Catalog}
We compiled a GBM transient event catalog with our search results.  Each candidate event was given an event ID with the naming convention that includes the search method initial letter (S, P, or B), mode (1, 2, 3, or 4), date (YYMMDD), hour (00$-$23), and event number during that hour; for example, \texttt{B1\_130112\_08\_1} is the event identified with the Bayesian method Mode 1, observed on 2013 January 12 in the 8th hour UT, the first event we identified in that hour.
Note that for the events before November 26, 2012, the event ID format has 3-digit fraction of day (000$-$999) instead of hour: for example, \texttt{B1\_100718244\_1}, corresponding to the CTTE datafile name.
The catalog consists of the following information for each event candidate: the event ID, event time (in \textit{Fermi} MET and in UTC), ``triggered detector" flag (12-digit binary, 0 or 1 for the 12 NaI detectors), detection significance for each detector, event duration, hardness ratio with $E_{\rm piv} =$ 25, 50, 75, and 300\,keV, class grades and flags (and the source name in case of a match), and the Bayesian probabilities for a set of event classes.
The full catalog is provided as 12 machine-readable tables and the sample entries are shown in \autoref{tab:catalog}.
The catalog is also accessible online in the accompanying Zenodo repository \citep{zenodo-catalog} and with the searchable database at \url{https://magnetars.sabanciuniv.edu/gbm}, where lightcurve plots are additionally provided for each candidate event: 12-detector lightcurve plot (for all events) and energy-resolved lightcurve of the brightest detector (only for SNR and Poisson events), the examples of which are shown in \autoref{fig:snr-example}, \autoref{fig:poi-example}, and \autoref{fig:bay-example}.

\begin{figure}[htbp!]
    \centering
    \includegraphics[width=0.9\textwidth, trim=60 200 40 230, clip]{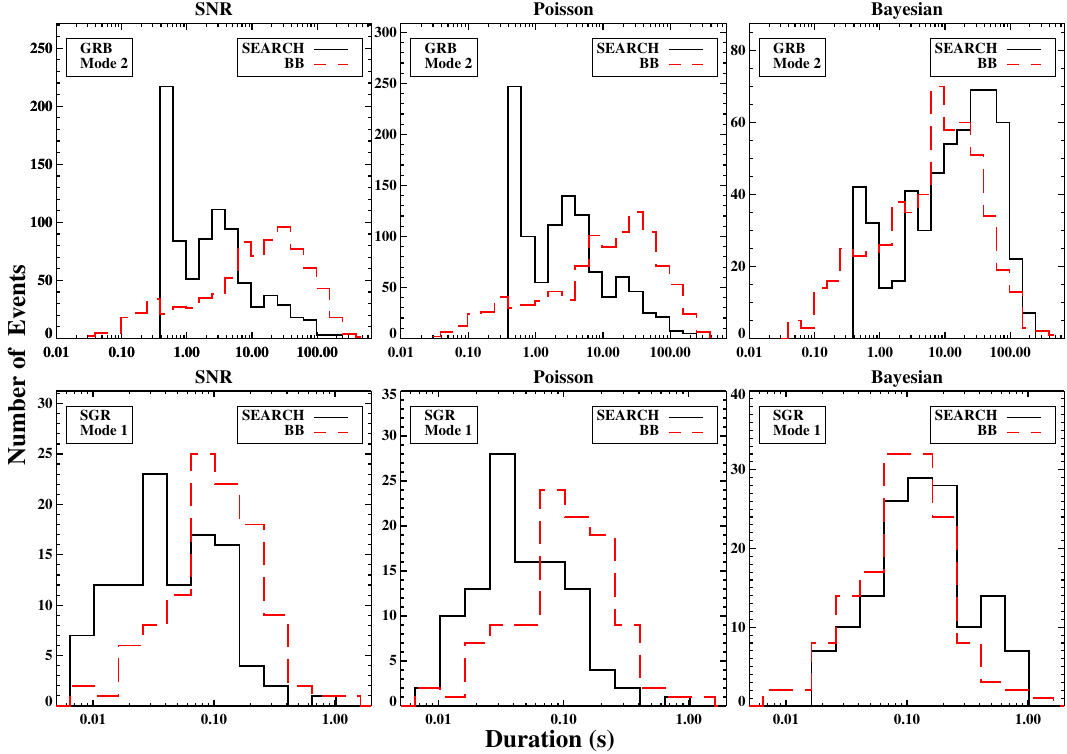}
    \caption{The comparison between the calculated BB duration and the duration included in the catalog for GBM triggered GRBs (top panels) and SGRs (bottom panels).  The BB duration was calculated in the energy ranges corresponding to Modes 2 and 1 for GRBs and SGRs, respectively.  By default, the minimum duration in our catalog is the data time resolution.
    }
    \label{fig:duration-comparison}
\end{figure}

As stated in \S\ref{subsec:untrig_classification}, we did not newly calculate the event duration included in the catalog; rather, it is the number of consecutive time bins with signal detection multiplied by the time resolution of the search mode (for the SNR and Poisson events) and the event block duration for the Bayesian events.  Using GBM-triggered events, when we compare this with the calculated duration using Bayesian-block lightcurves in the energy ranges appropriate for the event types (see \autoref{app1}), the duration presented in our catalog are mostly consistent; however, those estimated with the SNR and Poisson results tend to be shorter especially for events longer than a few seconds as seen in \autoref{fig:duration-comparison}. We note that the event duration for longer events included here were identified in our search multiple times as separate consecutive events.  In those cases, we combined their event duration values only for the purpose of this figure.
\begin{splitdeluxetable*}{lcccccccccccccccBc|cccc|cccc|ccc|ccccccccc}
\movetableright=-2.5in
\rotate
\tablecaption{The transient event catalog for the Bayesian Mode-1 search.  Some entries are skipped so as to display variations of possible entries. The catalog in its entirety is available as 12 machine-readable tables. 
    \label{tab:catalog}}
    \tabletypesize{\tiny}
    \tablewidth{0pt}
    \tablecolumns{37}
\tablehead{ & \multicolumn{2}{c}{Event Time} & \colhead{Triggered} & \multicolumn{12}{c}{Detection Significance$^a$} & \colhead{Duration} & \multicolumn{4}{c}{Hardness Ratio} & \multicolumn{4}{c}{Flag Grades$^b$} & \multicolumn{3}{c}{Flag IDs$^c$} & \multicolumn{9}{c}{Class Probabilities}\\
\cline{2-3} \cline{5-16}  \cline{18-37}\\
\colhead{Event ID} & \colhead{\textit{Fermi} MET} &	\colhead{UTC} &	\colhead{Detectors} & \colhead{N0} & \colhead{N1} &	\colhead{N2} & \colhead{N3} & \colhead{N4} & \colhead{N5} & \colhead{N6} & \colhead{N7} & \colhead{N8}	& \colhead{N9} & \colhead{Na} & \colhead{Nb} & \colhead{(s)} & \colhead{HR25} &	\colhead{HR50} & \colhead{HR75} & \colhead{HR300} & \colhead{TRIG} &	\colhead{SFL} & \colhead{TRST} & \colhead{PAR} & \colhead{TRIG}	& \colhead{SFL}	& \colhead{TRST} &	\colhead{SGR} &	\colhead{SGRB} & \colhead{LGRB} & \colhead{SFL}	& \colhead{TGF}	& \colhead{TRST} & \colhead{LPAR} & \colhead{DPAR} & \colhead{NO\_EVAL}\\
}
\colnumbers
\startdata
B1\_100716044\_1 & 300935620.299 & 2010-07-16T01:13:38.299 & 111111111111 & 9.15E+2 & 8.95E+2 & 1.13E+3 & 1.08E+3 & 7.16E+2 & 5.66E+2 & 1.04E+3 & 1.52E+3 & 7.46E+2 & 1.15E+3 & 1.39E+3 & 1.27E+3 & 0.232 & 1.190 & 0.696 & 0.430 & 0.220 & 0 & 0 & 0 & 1 & UN & NS & NT & 0.000 & 1.000 & 0.000 & 0.000 & 0.000 & 0.000 & 0.000 & 0.000 & 0.000 \\
B1\_100716113\_1 & 300941667.298 & 2010-07-16T02:54:25.298 & 111111111111 & 1.25E+3 & 1.33E+3 & 7.24E+2 & 1.50E+3 & 1.25E+3 & 9.56E+2 & 1.07E+3 & 1.18E+3 & 1.20E+3 & 9.98E+2 & 8.36E+2 & 9.60E+2 & 0.112 & 1.748 & 0.892 & 0.803 & 0.413 & 0 & 0	& 0	& 1	& UN & NS & NT & 0.000 & 1.000 & 0.000 & 0.000 & 0.000 & 0.000 & 0.000 & 0.000 & 0.000 \\
B1\_100717251\_1 & 301039705.900 & 2010-07-17T06:08:23.900
 & 111111111111 & 9.70E+2 & 8.09E+2 & 1.18E+3 & 1.14E+3 & 8.28E+2 & 6.84E+2 & 1.29E+3 & 1.22E+3 & 1.21E+3 & 1.23E+3 & 1.64E+3 & 1.17E+3 & 0.112 & 1.711 & 0.719 & 0.588 & 0.501 & 0 & 0 & 0 &1 & UN & NS & NT & 0.000 & 1.000 & 0.000 & 0.000 & 0.000 & 0.000 & 0.000 & 0.000 & 0.000 \\
... & ... & ... & ... & ... & ... & ... & ... & ... & ... & ... & ... & ... & ... & ... & ... & ... & ... & ... & ... & ... & ... & ... & ... & ... & ... & ... & ... & ... & ... & ... & ... & ... & ... & ... & ... & ... \\
B1\_110307838\_1 & 321221283.796 & 2011-03-07T20:08:01.796 & 010111000000 & 0.00E+0 & 1.26E+4 & 0.00E+0 & 1.02E+4 & 7.26E+3 & 1.72E+4 & 0.00E+0 & 0.00E+0 & 0.00E+0 & 0.00E+0 & 0.00E+0 & 0.00E+0 & 0.008 & 0.000 & 0.000 & 0.000 & 0.000 & 2 & 3 & 0 & 0 & SFLARE110307858 & 110307\_2002 & NT & 0.000 & 0.000 & 0.000 & 1.000 & 0.000 & 0.000 & 0.000 & 0.000 & 0.000 \\
... & ... & ... & ... & ... & ... & ... & ... & ... & ... & ... & ... & ... & ... & ... & ... & ... & ... & ... & ... & ... & ... & ... & ... & ... & ... & ... & ... & ... & ... & ... & ... & ... & ... & ... & ... & ... \\
B1\_130113\_03\_1 & 379741513.257 & 2013-01-13T03:45:10.257	& 011111000000 & 0.00E+0 & 2.41E+3	& 1.40E+3 & 3.52E+3	& 2.68E+3 & 5.95E+3 & 0.00E+0	& 0.00E+0	& 0.00E+0	& 0.00E+0 &	0.00E+0	& 0.00E+0 & 0.024 & 0.000 & 0.000 & 0.000 & 0.000 &	3	& 0	& 0	& 0	& SGR130113156 & NS & NT	& 1.000	& 0.000 & 0.000 & 0.000	& 0.000 & 0.000 &	0.000 &	0.000 &	0.000 \\
... & ... & ... & ... & ... & ... & ... & ... & ... & ... & ... & ... & ... & ... & ... & ... & ... & ... & ... & ... & ... & ... & ... & ... & ... & ... & ... & ... & ... & ... & ... & ... & ... & ... & ... & ... & ... \\ 
B1\_150621\_08\_9 & 456567216.712 & 2015-06-21T08:13:33.712 & 000000011000	& 0.00E+0 &	0.00E+0 & 0.00E+0 & 0.00E+0 & 0.00E+0 & 0.00E+0	& 0.00E+0 & 1.19E+3 & 1.28E+3 & 0.00E+0 & 0.00E+0 & 0.00E+0 & 1.441 &	0.000 & 0.000 &	0.000 & 0.000 & 0	& 0 & 3 & 0 & UN & NS & V404-CYG & 0.000 & 0.000 & 0.000 & 0.000 & 0.000 & 1.000 & 0.000 & 0.000 & 0.000 \\
... & ... & ... & ... & ... & ... & ... & ... & ... & ... & ... & ... & ... & ... & ... & ... & ... & ... & ... & ... & ... & ... & ... & ... & ... & ... & ... & ... & ... & ... & ... & ... & ... & ... & ... & ... & ... \\
B1\_170908\_17\_52 & 526585480.187 & 2017-09-08T17:44:35.187 & 111111111111 & 7.07E+2 & 6.26E+2 & 8.17E+2 & 2.80E+3 & 2.66E+3 & 1.32E+3 & 2.94E+3 & 4.06E+3 & 1.84E+3 & 1.19E+3 & 2.99E+3 & 1.07E+3 & 4.681 & 0.000 & 0.000 & 0.000 & 0.000 & 3 & 0 & 0 & 1 & LOCLPAR170908739 & NS & NT & 0.000 & 0.000 & 0.000 & 0.000 & 0.000 & 0.000 & 1.000 & 0.000 & 0.000 \\ 
... & ... & ... & ... & ... & ... & ... & ... & ... & ... & ... & ... & ... & ... & ... & ... & ... & ... & ... & ... & ... & ... & ... & ... & ... & ... & ... & ... & ... & ... & ... & ... & ... & ... & ... & ... & ... \\
B1\_210106\_22\_2 & 631664560.808 & 2021-01-06T22:22:35.808	& 000001100000 & 0.00E+0 & 0.00E+0	& 0.00E+0 &	0.00E+0 & 0.00E+0 & 3.39E+2 & 9.57E+2 & 0.00E+0 & 0.00E+0 & 0.00E+0 & 0.00E+0 & 0.00E+0 & 0.265 & 1.159 & 0.472 & 0.307 & 0.433	& 0	& 0	& 0	& 0	& UN & NS & NT & 0.784 & 0.216	& 0.000 & 0.000 & 0.000 & 0.000 & 0.000 & 0.000 & 0.000\\
\hline
\hline
\enddata
\vspace{0.2cm}
\tablecomments{ $^a$ SNR (in the unit of $\sigma$) for SNR search, Poisson probability for Poisson search, and peak count rates (in count/s) for Bayesian search\\
$^b$ The assigned grades (see \S\ref{subsec:filtering}) for GBM triggers (TRIG), solar flares (SFL) and X-ray transient sources (TRST), corresponding to the Flag ID listed in the following column. Only 0 or 1 for particle events solely based on the spacecraft position at the time of the event, so these events also have associated Class Probabilities. \\
$^c$ TRIG = Untriggered (UN) or \textit{Fermi} trigger ID if matches (grade >0), SFL = Non-SFLARE (NS) or SFLARE ID from the FERMIGSOL catalog if matches, TRST = Non-Transient (NT)
or Transient Name if matches}
\tablecomments{The catalog is also available as CSV files on the Zenodo repository \citep{zenodo-catalog}}
\end{splitdeluxetable*}

\subsection{Known-Event Flagging Results}
In this section, we present the statistics and the breakdown of the known-event flagging results, namely for GBM triggered events, solar flares,  active X-ray sources, and particle events.

\textbf{GBM Triggered Events:} In \autoref{tab:trig_stat}, we summarize the number of triggered events we identified in our search, sorted by the event classes.  As seen in the table, not all GBM triggered events were identified in our search, with the highest identification percentage (90\%) associated with the triggered SGR events. This was expected due to the difference in the detection criteria used here; however, the detection rates of the triggered events were unexpectedly low.  While investigating the possible reasons for non-identifications, we noticed that 97\% of undetected triggers that occurred before November 26, 2012 fell into the data gaps of CTTE data provided by the GBM team.  Note that the date is when the CTTE data format changed to an hourly format (i.e., 24 hourly CTTE files a day).  Moreover, most of the rest (3\%) occurred at the very last second of the corresponding dataset, which hinders the detection of the triggered events. 
The CTTE data before this date were accumulated until a trigger happens, at which point the CTTE data accumulation halts and the trigger-mode TTE data accumulation starts.  The accumulation of CTTE data resumes after the trigger mode ends (in 300$-$600\,s).  Since our searches did not utilize the trigger-mode TTE data, the triggered event data were naturally missing.  This is clearly seen in \autoref{tab:trig_stat}.
For triggered events for which the CTTE data exist, other possible reasons for non-identifications include differences in the background estimation methods and in the triggering criteria (GBM flight software employs 119 trigger algorithms, including BGO detector criteria) in terms of energy ranges and sensitivities. Many of the non-identified triggered events were flagged in single detectors in our search, with the second brightest detectors' count rates falling short of our trigger criteria and thus, were not identified as events in our search. The statistics presented in the table also include the summary of the assigned grades 3, 2, and 1 (see \S\ref{subsec:filtering}), with 3 being the most definitive match.  It can be seen that $\gtrsim 90$\% of the events were matched to the triggers with high confidence with Grade 3.  For SGRs, 13 out of 19 undetected triggers have durations shorter than 8\,ms, which is the minimum time resolution of our searches.

\begin{table}[!htbp]
    \centering
    \caption{The number of GBM triggered events identified in the search (all modes and methods combined) and the summary of the assigned grades}
    \label{tab:trig_stat}
    \begin{tabular}{|c|c|c|c|c|c|c|c|}	
    \hline											
\textbf{Event}	 & 	\textbf{Total No. of}	 & \multicolumn{3}{c|}{\textbf{Identified (\%)}} & 	\multicolumn{3}{c|}{\textbf{Grades}}					\\ \cline{3-8}
\textbf{Class}	 & 	\textbf{Triggered Events}*	 & 	\textbf{Pre-24h}**	 & 	\textbf{Post-24h}	 & \textbf{All} &
\textbf{3}	 & 	\textbf{2}	 & 	\textbf{1}	\\ \hline \hline
GRB	 & 	2588	 & 7 (1\%)& 1858 (91\%) &	1865 (72\%)		 & 	1574	 & 	257	 & 	34	\\
SGR	 & 	193	 & 0 (0\%) & 174 (99\%) &	174 (90\%)		 & 	174	 & 	0	 & 	0	\\
SFLARE	 & 	1144 	 & 97 (24\%) & 689 (92\%) &	786	(69\%) 	 & 	679	 & 	81	 & 	26	\\
TGF	 & 	579	 & 2 (2\%) & 202 (45\%) &	204	(35\%)  	 & 	92	 & 	76	 & 	36	\\
\hline	   
    \end{tabular}
\tablecomments{* The number of GBM triggered events during the search period.  For TGFs, the number of events excludes the ones triggered solely by BGO detectors (GBM algorithm number 119; \citealt{vonkienlin2020}). \\
** Before and after November 26, 2012, on which the 24-hour format CTTE became available}
\end{table}

In addition, we present in \autoref{tab:trig_stat_mode} the breakdown of the identification percentages in each search mode.  The mode in which each event class was identified with the highest percentages are indicated in bold, which match our expectations by the definition of the four modes (\autoref{tab:modes}).
More specifically, Mode-1 search was done in the shortest time resolution (8\,ms) in the energy $<100$\,keV, fit to the characteristics of SGR bursts; Modes 2 and 3 are with a bit longer time resolutions and encompasses higher energy, which is more suitable for GRBs and TGFs. Here, we use the definition of short GRBs as those with published $T_{90}$ duration $\leq$ 2.0\,s.  Finally, we identified 54\% of the triggered SFLAREs equally in Modes 1, 2, and 4, since SFLAREs emit most in softer energy below 20\,keV \citep{Pesce-Rollins_2015}.
\begin{table}[!htbp]
    \centering
    \caption{The breakdown by the search Modes of the percentages of GBM triggered events found in our search.  The Mode with the highest identification percentage in each class is shown as bold.}
    \label{tab:trig_stat_mode}
    \begin{tabular}{|l|c|c|c|c|c|}
    \hline
	\textbf{Event}&  	\textbf{Total No. of} &	\multicolumn{4}{c|}{\textbf{Percentage (\%)}} \\
 \cline{3-6}
 \textbf{Class}& 	\textbf{Triggered Events} &	Mode 1	&	Mode 2	&	Mode 3	&	Mode 4	\\ \hline\hline
GRB\tablenotemark{a}	& 	2588 &	24	&	\textbf{68}	&	34	&	42	\\
\hspace{0.5cm}SGRB	&  418 &	37	&	61	&	\textbf{66}	&	16	\\
\hspace{0.5cm}LGRB	& 2170 &	21	&	\textbf{70}	&	28	&	47	\\
SGR	& 	193	 &	\textbf{90}	&	82	&	45	&	64	\\
SFLARE	& 	1144	&	\textbf{54}	&	\textbf{54}	&	26	&	\textbf{54}	\\
TGF	& 	579	&	5	&	3	&	\textbf{32}	&	6	\\
    \hline
\end{tabular}
\\[0.3pt]
\tablenotetext{a}{SGRB/LGRB = Short/Long GRB with published $T_{90}$ duration less than / greater than 2.0\,s }
\tablecomments{Events identified in multiple modes are counted in each of the modes, so there are duplicates.\\
Some long events can trigger the detectors multiple times within their duration, so the number of triggered events in our search results can be more than the actual number of events presented here.} 
\end{table}

\textbf{Untriggered SFLAREs:} Within our search period, there are 5110 SFLAREs reported in the \textit{Fermi}-GBM Solar Flare Catalog, which includes both 976 GBM triggered flares (responsible for 1144 triggers, since some flares triggered GBM multiple times within the event duration) and flares that did not trigger GBM. Overall, we found 3786 (74\%) of them in our search, mostly with the Poisson and SNR methods (Grade 3), accounting for a total of 182,555 events in all modes and methods.  They were detected with Mode 1 mostly in all methods (see \autoref{tab:transnt_stat_mode}). Note that some of these multiple detections within a single SFLARE could be instrumental; SFLAREs can be extremely bright that the count rates can exceed the instrument limit of 375\,kHz when all 14 GBM detectors data are summed \citep{meegan2009}.  Bright parts of SFLAREs that exceed this limit saturate and the binned CTTE data could show data spikes and drops, which are detected in our search.  All of these are identified as SFLAREs with the grade of 3.

\textbf{Known X-ray Source Activities:} We identified candidate events that are due to the signals from four active X-ray sources: V404 CYG (was active in June 2015), Swift J0243.6+6124 (active in November 2017), MAXI J1820+070 (in April 2018) and A0535+262 (in November 2020). These four sources triggered the GBM detectors a total of 414 times during their active periods. 
Our search identified a total of 420,335 triggered and untriggered signals from these sources (Grade 3 in all modes and methods), with the majority found in Modes 2 and 4 of the SNR and Poisson methods, which are more susceptible to multiple detections from single spiky, long events than the Bayesian method. (see  \autoref{tab:transnt_stat_mode}).
Many of these activities show periodic pulsation behaviors.
Overall, about a half of the events found in our all modes and methods are due to SFLAREs and X-ray transient sources. 

\begin{table}[!htbp]
    \centering
    \caption{The number of events identified as TRANSNTs and SFLAREs with Grade 3}
    \label{tab:transnt_stat_mode}
    \begin{tabular}{|l|c|c|c|c||c|}
    \hline

 \textbf{Method} &	\textbf{Mode 1}	&	\textbf{Mode 2}	&	\textbf{Mode 3}	&	\textbf{Mode 4} & 	\textbf{All}	\\ \hline 

  	\multicolumn{6}{|c|}{\textbf{Number of Transient Detections (\%)}}   \\ \hline
\textbf{SNR}	& 	3,458 (2.7) &	80,490 (62.7)	&	48 (0.08)	&	81,281 (63.3)	&	165,277	\\
\textbf{Poisson}	& 	3,942 (3.2)	 &	124,982 (46.3)	&	42 (0.09)	& 112,509 (42.1)	&	241,475	\\
\textbf{Bayesian}	& 	11,307 (13.3)	&	1,869 (9.5)	&	227 (0.94)	&	180 (1)	& 13,583	\\
    \hline
    \hline
\textbf{All}	& 	18,707	&	207,341	&	317	& 193,970	& 420,335	\\
    \hline
     	\multicolumn{6}{|c|}{\textbf{Number of SFLARE Detections (\%)}}   \\
   \hline 
 
\textbf{SNR}	& 	34,226 (26.8) &	17,205 (13.4)	&	6,254 (9.8)	&	9,155 (7.1)	&	66,840	\\
\textbf{Poisson}	& 	51,075 (41.4)	 &	23,365 (8.7)	&	6,491 (13.3)	& 12,406 (4.6)	&	93,337	\\
\textbf{Bayesian}	& 	19,822 (23.4)	&	37 (0.2)	&	2,485 (10.3)	&	34 (0.2)	& 22,378	\\
    \hline
    \hline
\textbf{All}	& 	105,123	&	40,607	&	15,230	& 21,595	& 182,555	\\
    \hline
\end{tabular}
\\[0.3pt]
\tablecomments{The numbers indicated in parentheses are percentages.} 
\end{table}

Moreover, we also checked for potential detections due to occultation steps of bright X-ray sources, using one of the brightest sources, CYG-X1 out of the $\sim$250 X-ray sources listed by the GBM team\footnote{\url{https://gammaray.msfc.nasa.gov/gbm/science/earth_occ.html}}.  To identify an occultation step among our detected events, our criterion was whether any rise time of the source falls within the range of $\pm$5 s of our detection times. If so, we checked whether at least 2 out of 3 brightest detectors see the source with an angle of less than 60$^{\rm o}$. Using CYG-X1, we did not identify any candidate events potentially due to the occultation steps.  We then concluded that the probability of detecting occultation steps due to other dimmer X-ray sources in our searches is negligible.  Additionally, persistent emission flux levels often increase when an X-ray source become active, causing the corresponding occultation steps to be more prominent.  Therefore, we also checked whether we detect any occultation steps due to the four sources during their active periods by applying the same procedure above. We only have a few occultation step detections for the four sources in all methods and modes. In these rare cases, however, we had already associated the events with these sources with a grade of 3 since the source locations matched.

\textbf{Particle Events:} Following our definition of particle event flags based on the spacecraft positions (see \S\ref{subsec:filtering}), about one third of all event candidates were flagged as potential particle events (flag = 1) in all search modes (on average, 30.5\% for the SNR searches, 31.5\% for the Poisson searches and 35.1\% for the Bayesian searches).  Note that the particle event flag of 1 only indicates an elevated probability of the event being due to charged particles, and does not mean definitive classification.

\vspace{12pt}
The resulting total numbers of known sources that we flagged are summarized in \autoref{tab:filtering-summary}, along with the numbers of remaining events with ``unknown" sources or classifications, which
were subjected to the event classification algorithm described in \S\ref{subsec:untrig_classification}. 
Here, we refer to the events that were matched with grade 3 (either with GBM triggers, SFLAREs, or TRANSNTs) as ``known" events, and the rest were subjected to classification.  Therefore, they are included as ``unknown" events in \autoref{tab:filtering-summary}. 

\begin{table}[!htbp]
    \centering
    \caption{The numbers of events identified in our search matched with known sources (``Known" columns) and the remaining (``Unknown") events subjected to the classification algorithm.  The time resolution and the energy range used in each mode are also shown as a reference.}
    \label{tab:filtering-summary}
    \hspace*{-2.5cm}
    \begin{tabular}{|cccccccccccc|}
    \hline
	&	\multicolumn{2}{c}{\textbf{Mode 1}}			&&	\multicolumn{2}{c}{\textbf{Mode 2}}			&&	\multicolumn{2}{c}{\textbf{Mode 3}} &&	\multicolumn{2}{c|}{\textbf{Mode 4}} \\
    &	\multicolumn{2}{c}{[8\,ms, 10--100\,keV]}			&&	\multicolumn{2}{c}{[512\,ms, 10--300\,keV]}	&&	\multicolumn{2}{c}{[16\,ms, 50--1000\,keV]}			&&	\multicolumn{2}{c|}{[2\,s, 10--25\,keV]}			\\[2px]	\cline{2-3} \cline{5-6} \cline{8-9} \cline{11-12}						
\textbf{Method}	&	Known	&	Unknown	&&	Known	&	Unknown	&&	Known	&	Unknown	&&	Known	&	Unknown	\\	\hline	\hline					
SNR	&	42,627	&	85,277	&&	100,357	& 27,945	&&	9,748	&	54,165	&&	91,446	&	37,065	\\							
Poisson	&	60,397	&	63,048	&&	152,772	&	116,917	&&	10,049	&	38,661	&&	126,382	&	140,849	\\							
Bayesian	&	35,654	&	48,972	&&	2,697	&	17,047	&&	4,626	&	19,387	&&	565	& 17,470	\\							
\hline
\end{tabular}
\tablecomments{``Unknown" events also include those matched to known sources with grades lower than 3, indicating uncertain association.}
\end{table}

\subsubsection{Unknown Event Classification Results}
We applied the unknown event classification algorithm with Bayesian probability based on their spectral hardness and duration (described in \S\ref{subsec:untrig_classification}) to all the ``Unknown" events in \autoref{tab:filtering-summary}, separately for each search mode and method.
The resulting probabilities are listed in the catalog (\autoref{tab:catalog}) for all events, both known and unknown, for the event classes of SGR, SGRB, LGRB, SFLARE, TGF, TRANSNT, LOCLPAR, and DISTPAR.  In the cases where the event class cannot be assigned, the NO\_EVALUATION was set to 1.  Known events are assigned the probability of 1 to the known event class.

Furthermore, for events identified (with high probability) as SGR or TRANSNT, it is possible to identify the specific source that was active at the time of the detected events.  To identify and confirm the association with a particular source, however, a more accurate localization of events beyond the method based only on the 12-detector count rate comparison is crucial.  There are many potential sources that became active during our search period, and thus such a localization effort is not trivial and requires extensive additional analysis.  Besides, the localization accuracy provided solely by \fermi is limited, with a minimum uncertainty of a few degrees. 
Nonetheless, as we demonstrated in \citet{uzuner2023} for two sources (Swift\,J1818.0-1607 and PSR\,J1846.4-0258), it is possible to sufficiently localize events using “\textit{Targeted Search}” algorithm mentioned in \S\ref{sec:intro} developed by the GBM team.  We have not performed the “\textit{Targeted Search}” localization for other sources since this is out of scope of the search project.  In the catalog, source names are included for events with confirmed or potential association to a particular source.

\section{Conclusion} \label{sec:discussion}
We obtained the large database of short transient events in the \fermi data of 11 years, via the comprehensive offline searches with four modes and three statistical methods.
Our results expand the currently-existing list of known events by about two-folds overall.
The vast majority of the detected event samples of the 12 searches overlap. However, since each search mode and method were more effective at detecting particular types of events, each of the 12 searches includes a unique set of events that were only detected in the particular mode and method.
Note that the percentages of the overlaps are not determined here;
This is because the exact time of the detection may be slightly different in various modes and methods even for the same transient event and the detectors with highest counts may also vary slightly especially for weaker events. In addition, the total duration of the events, especially for multi-episodic events, cannot be confirmed without further investigations.  The aim of this paper is to present the search methods and results along with the potential classifications.

We present the resulting \fermi transient event catalog that is accessible online\footnote{Available as machine-readable tables as well as at the accompanying digital repository on Zenodo \citep{zenodo-catalog} and on our own server \url{https://magnetars.sabanciuniv.edu} with a searchable, most up-to-date database}.
Using our catalog, users can select the search method and mode that are most suitable for their astrophysical objectives, to uncover previously undetected transient events of their interests.
One note for our search using the Bayesian block method is that the initial ``source" detection time interval is set to $\leq$\,1 for Modes 1 and 3, which by default causes the detected events to be of the order or shorter than 1 s unless the consecutive short time bins are detected.
This was done to optimize the detections of short GRBs and SGR bursts.  Indeed, the majority of the events found in Mode 1 and Mode 3 of Bayesian block search (79\% and 89\%, respectively) are events with the duration $\lesssim$\,1\,s.  This, however, does not mean that longer-duration events cannot be identified in the Bayesian block search since we used longer event time of 100 and 200 s in Modes 2 and 4, respectively; in fact, the mean duration of the events found in Modes 2 and 4 are 92\,s and 214\,s respectively, with more than 95\% of the events are longer than 1\,s.

Looking at the search results (\autoref{tab:search_summary}), the number of events detected with Bayesian methods in Modes 1 and 3 are comparable to the SNR and Poisson results.  However, overall the Bayesian detections are much less than those of the SNR or Poisson searches.  The differences in the detection numbers can be attributed to the following:
The Poisson and SNR methods turned out to be more sensitive to short dim events compared to the Bayesian method; the Bayesian method detects brighter events.  This is evidenced by the fact that while attempting to calculate the duration of events identified only with SNR or Poisson, many short dim events were not appearing as a separate block in the Bayesian-block representation of the corresponding lightcurves, as stated in \S\ref{subsec:untrig_classification}.  Furthermore, we observed that the SNR and Poisson methods are more prone to detecting a single event multiple times, contributing to higher number of detections.  This is especially the case for spiky, highly-variable, long events such as SFLAREs.  As can be estimated from \autoref{tab:trig_stat_mode} and \autoref{tab:transnt_stat_mode}, the SNR and Poisson methods detects 10-80 times per SFLARE on average in all modes, whereas the Bayesian method detects far less SFLAREs overall and less detections per SFLARE, especially in Modes 2 and 4.

We also see very high rates of SAA events detected in the Bayesian search Modes 2 and 4.  The differences in the numbers (\%) of SAA events between the Bayesian method and the other two stem from the slowly varying lightcurves at the beginning and end of a data segment (when the detectors are turned on after or turned off before a SAA passage); they usually vary gradually with the time scales of $\sim$10 to 100 s.  Since the widths of Bayesian blocks are often longer than the time resolution of the data for each search mode, which the SNR and Poisson methods are based on, the Bayesian methods can identify longer blocks as events.  This effect is more evident in Modes 2 and 4, where the maximum duration of a potential event is set high (100 and 200 s, respectively).  In fact, in these modes, almost all SAA-filtered events are of long durations ($>$10-1000 s).

Building on this work, we plan to expand our searches to include more recent data (July 2021 onward, possibly for the next decade of GBM observation) and enhance our unknown-event classification algorithm.  For the latter, we intend to incorporate machine-learning algorithms, trained with a comprehensive dataset including both triggered events and untriggered events found in existing literature. 
It is planned that the updated results will be shared publicly through our website with a searchable interface (\url{https://magnetars.sabanciuniv.edu}) as they become available.

\section*{acknowledgments}

We thank the anonymous referee for their insightful comments and refinement suggestions that significantly improved the quality of this paper. We also thank Michael S. Briggs and Colleen Wilson-Hodge for their support for the clarification of the GBM classification algorithms, data issues and X-ray source observations.
The authors acknowledge the support from the Scientific and Technological Research Council of Turkey (T\"{U}B\.{I}TAK grant No.\,118F344), and the use of the \textit{Fermi} Solar Flare Observations facility funded by the \textit{Fermi} GI program (\url{http://hesperia.gsfc.nasa.gov/fermi_solar/}).

\appendix
\section{Characterization of Triggered Events} \label{app1} 
In this pre-study of collective properties of all classified triggered events, we calculated and compared two properties: Hardness Ratio (HR) and duration distributions, for each of these event classes. We used all triggered events until the end of 2020, excluding LOCLPAR, DISTPAR, and TRANSNT events, for this study.

\begin{itemize}

\item \textbf{Hardness Ratio Comparisons:} Hardness ratios (HRs) are used as a quick measure of the general hardness of a population of a certain type of astrophysical transients. They can be determined in terms of observed counts as well as in terms of deconvolved photon counts, inferred based on the photon model (of user’s choice) that best fits the observed spectrum. In the gamma-ray energy range in which the GBM operates, each incoming photon is redistributed to a wider spectrum after interacting with the detector, thus the observed count spectrum does not represent the original source spectrum. However, the HRs in observed count space could still be used as a rough estimate of the spectral hardness, without requiring additional processing and analysis of the raw data. We, therefore, studied the HRs of all types of triggered events using their background-subtracted observed counts in the energy range of 10–2000 keV, with three energy ``pivot" points ($E_{\rm piv}$): 25, 50 and 75 keV. The HR was defined as the ratio of background-subtracted counts in $E_{\rm piv}$–2000 keV (hard band) over 10 keV–$E_{\rm piv}$ (soft band), at the event peak time. These $E_{\rm piv}$ values were chosen since we found these hardness ratios to best distinguish events from different classes, e.g., SGR bursts from short GRBs and TGFs, all of which have similar event duration based on the triggered event properties.

Using this definition, we calculated the HRs with three $E_{\rm piv}$ values for GRBs, SFLAREs, SGRs, and TGFs together.  The procedure was as follows. We first estimated the background level by taking an average of the observed count rates in the interval from 17 seconds to 4 seconds before the trigger time of the event, using triggered TTE data binned to 8 ms. The background was estimated separately for the soft and hard energy bands. After subtracting the estimated background rate, the total source counts in the hard and soft energy bands were obtained for each time bin, and the associated uncertainty was calculated by propagating the errors in the observed count rates. To ensure that we had sufficient signals in each time bin to constrain the HRs, we combined the consecutive time bins until we obtained the HR value with its uncertainty being at least 3 times smaller ($\lesssim$ 30\%) than the HR value itself. The resulting time resolution varied between 8 ms to a few seconds, depending on the duration and brightness of the event. We then looked at these well-constrained HRs at the onset of each trigger as well as at the peak of each event, where the counts are the highest.
We note that, although we calculated the HR values at the trigger times of these events, we only used the peak HR values for the unknown event identification process described in \S\ref{subsec:untrig_classification}.  This is because the GBM trigger times often do not match exactly to the corresponding event-detection times in our search due to the differences in our detection algorithms and theirs (see \S\ref{subsec:filtering}, \textbf{GBM Triggered Events} description).  The overall distributions of trigger time HR and peak HR values do not significantly differ.

Once we calculated the onset HRs and peak HRs with three $E_{\rm piv}$ values for all event types, we compared the HR distributions of event class pairs of potentially similar properties, such as SFLAREs \& LGRBs (duration $>$ 2\,s) or SGRs \& SGRBs (duration $<$ 2\,s). The purpose is to find the most distinguishing HR parameters between each of these class pairs so that these distributions can be utilized as a priori probability functions to calculate the event class posterior likelihood. We used different $E_{\rm piv}$ for each comparison since it is not possible to classify all types of events with a single HR with the same $E_{\rm piv}$. 
In \autoref{fig:hr-comparison}, we show the peak HR distribution comparisons for SFLARE vs.~LGRB (with $E_{\rm piv} = 50$\,keV) and SGR vs.~SGRB (with $E_{\rm piv} = 75$\,keV), with the best-distinguishing (i.e., least overlapping) $E_{\rm piv}$ values.
We then fit log-normal functions to the peak HR distributions and obtained a continuous probability function representing each event class.
\begin{figure}[htbp]
    \centering
    \includegraphics[width=0.48 \textwidth,  trim=65 225 85 150, clip]{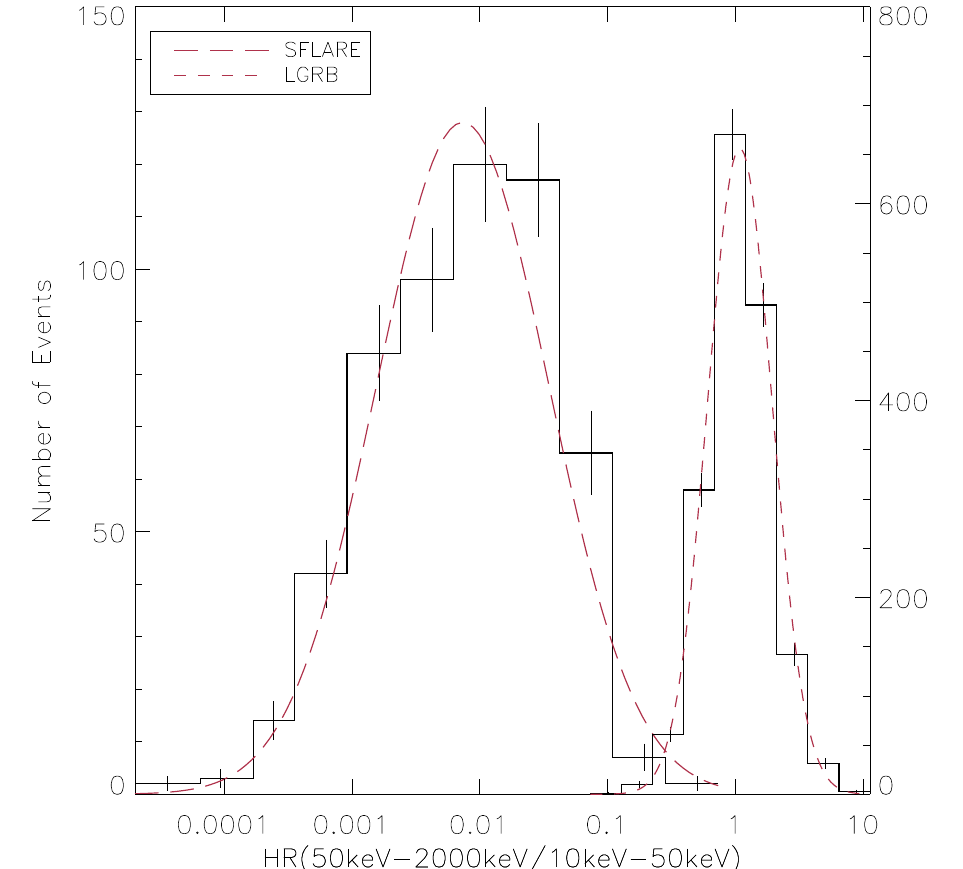}
    \includegraphics[width=0.48 \textwidth,  trim=65 225 85 150, clip]{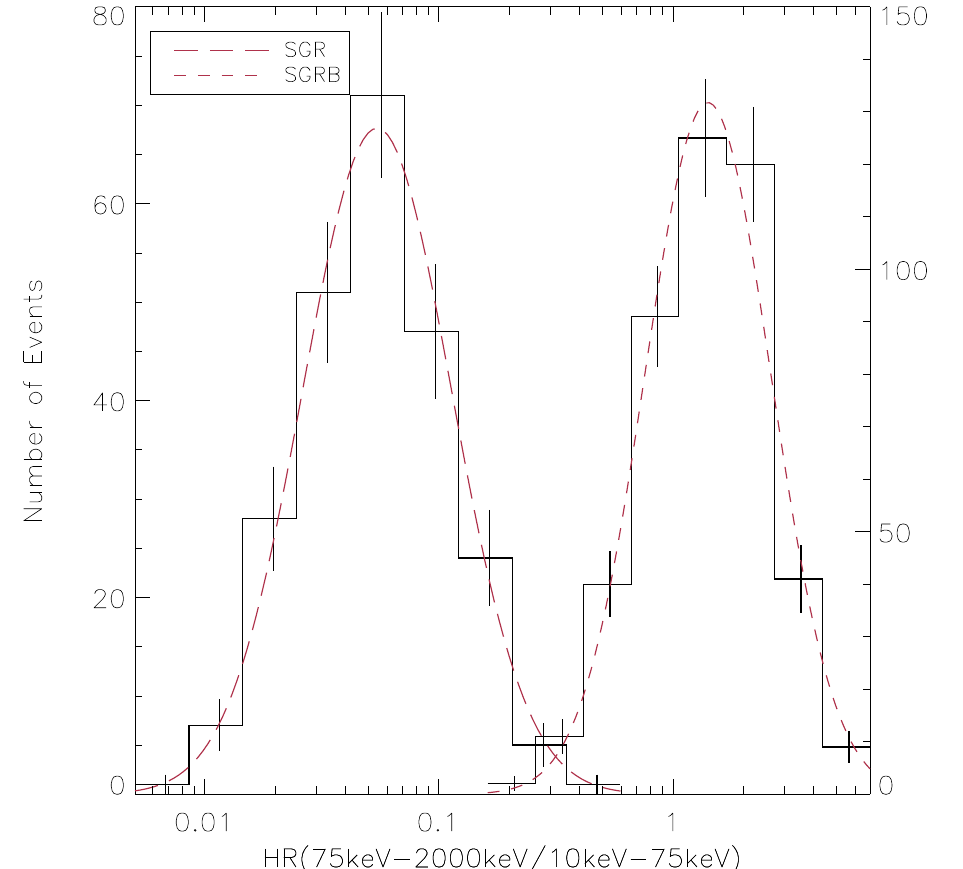}
    \caption{Distribution comparisons of HRs at the peak time between SFLAREs \& long GRBs ($>$2 s) [left panel] and SGRs \& short GRBs ($<$2 s) [right panel].  The dashed curves show log-normal fits to the distributions.}
    \label{fig:hr-comparison}
\end{figure}

\item \textbf{Duration Comparisons:} Another parameter that could potentially distinguish event classes is event duration.  Since the duration calculations require post-trigger analysis, the GBM trigger catalog does not include duration information.  However, for some event types such as GRBs and SFLAREs, the triggered event duration have been published in their own catalogs in standard energy ranges.  Here, we independently determined the event duration for all triggered events, using the Bayesian-Block (BB) representation of its lightcurve in various energy ranges specific to event classes: in 10--100\,keV for SGRs, 25--300\,keV for GRBs, 10--25\,keV for SFLAREs, and 300--1000\,keV for TGFs. To define the event time blocks (i.e., event duration), we need to first identify background blocks in the BB lightcurve.  
We selected a pre-defined minimum background block width for each type of event, based on their typical event durations. The minimum background widths used are 4\,s for SGR, 0.5\,s for TGFs, 60\,s for SFLAREs, $T_{90}$ for GRBs, where $T_{90}$ is the duration reported in the \textit{Fermi} GBM Burst Catalog \citep{vonkienlin2020} during which 90\% of all photons (in $50-300$\,keV) from a burst was detected. Any blocks in between the background blocks before and after the event trigger time, whose count rates are higher than the background count rates are labeled as event blocks. The BB duration, then, is the total width of the event blocks. 
The comparisons of the calculated BB duration and the published duration for GRBs and SFLAREs are shown in \autoref{fig:duration}.  
We confirmed that the BB duration values are consistent with $T_{90}$ for GRBs due to the very similar energy ranges used for the calculations.  The published SFLARE duration takes into account all energy $>$10\,keV while our calculations are done in 10--25\,keV.
We used the best-fit log-normal functions of the BB duration distributions of triggered events (each class separately) as parent distributions for the likelihood test for event classifications.
As an example, \autoref{fig:grb_dur_dist} shows the BB duration distribution of \ferminosp-triggered GRBs.
The fact that the calculated BB duration values are consistent with the values published in the catalogs (see \autoref{fig:duration}), validates that the same approach can be used to determine the durations of untriggered events (of unknown sources) found in our search. We note, however, that the BB approach to defining the event duration proved ineffective for many weak events found in our SNR or Poisson searches, as stated in \S\ref{subsec:untrig_classification}, as well as distinguishing event blocks and background blocks without pre-knowledge of event class posed challenges.  Therefore, for the events we found in the searches and included in the catalog, we did not newly estimate the individual event duration using this BB approach (see \S\ref{subsec:untrig_classification} for our event duration definition).

\end{itemize}
\begin{figure}[htbp]
    \centering
    \includegraphics[width=0.48 \textwidth,  trim=75 80 40 235, clip]{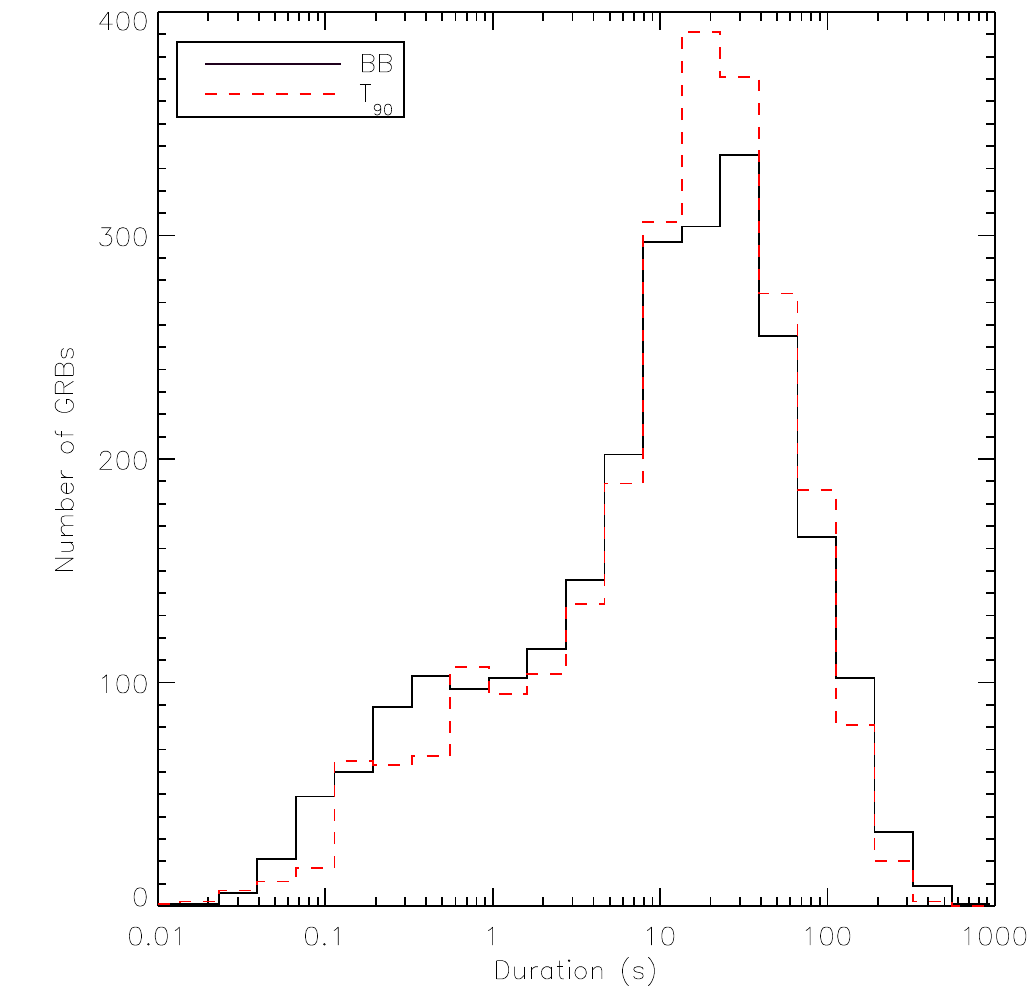}
    \includegraphics[width=0.48 \textwidth,  trim=75 80 40 235, clip]{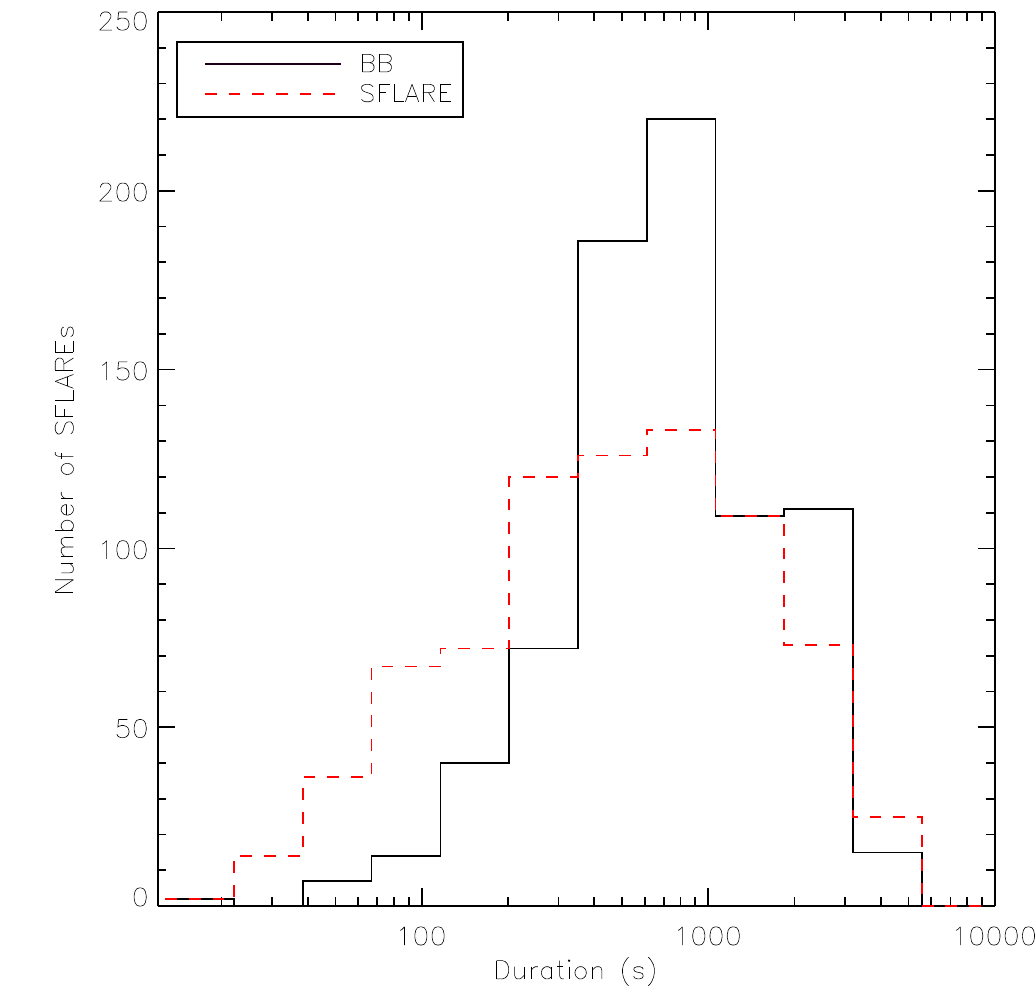}
    \caption{Distribution comparisons of BB durations vs.~the catalog values (red dashed lines) for the GBM-triggered GRBs [left panel] and SFLAREs [right panel].}
    \label{fig:duration}
\end{figure}
\begin{figure}
    \centering
    \includegraphics[width=0.55\linewidth, trim=30 300 25 50, clip]{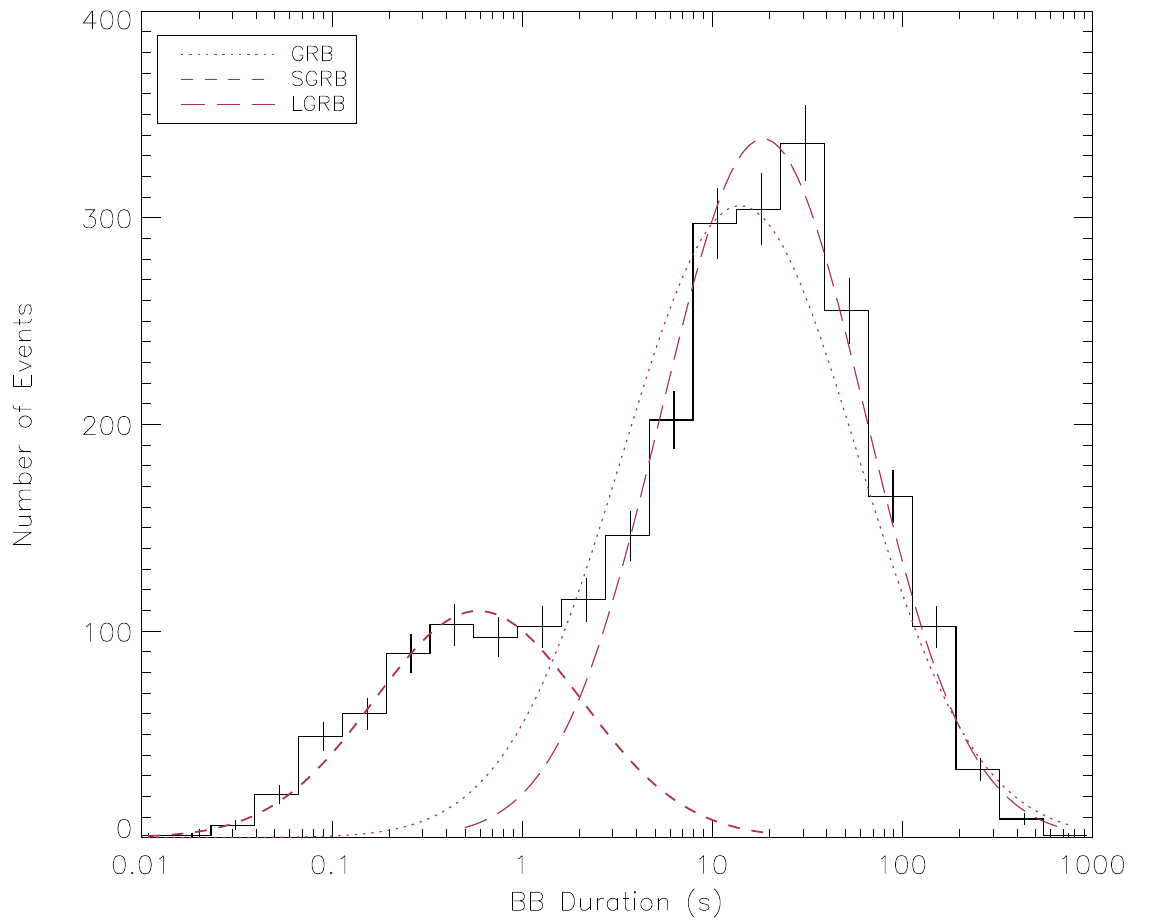}
    \caption{BB duration distribution of \fermi triggered GRBs in the 25-300 keV range. Long-dashed line indicates LGRB lognormal fit whereas short-dashed line indicates SGRB lognormal fit. Their intersection is $\sim$ 2 seconds. Also, the dotted line represents the overall GRB duration distribution.}
    \label{fig:grb_dur_dist}
\end{figure}

\bibliographystyle{aasjournal}
\bibliography{refs}

\end{document}